\documentclass[letterpaper]{jpconf}
\usepackage{graphicx}

\begin{document}

\title{The Sanford Underground Research Facility at Homestake}


\author{J Heise}

\address{Sanford Underground Research Facility, 630 East Summit Street, Lead, SD 57754}

\ead{jaret@sanfordlab.org}

\begin{abstract} 
The former Homestake gold mine in Lead, South Dakota has been transformed 
into a dedicated facility to pursue underground research in rare-process 
physics, as well as offering research opportunities in other disciplines 
such as biology, geology and engineering.  A key component of the Sanford 
Underground Research Facility (SURF) is the Davis Campus, which is in 
operation at the 4850-foot level (4300 m.w.e.) and currently hosts two 
main physics projects: the LUX dark matter experiment and the {\sc 
Majorana Demonstrator} neutrinoless double-beta decay experiment.  In 
addition, two low-background counters currently operate at the Davis 
Campus in support of current and future experiments.  Expansion of the 
underground laboratory space is underway at the 4850L Ross Campus in order 
to maintain and enhance low-background assay capabilities as well as to 
host a unique nuclear astrophysics accelerator facility.  Plans to 
accommodate other future experiments at SURF are also underway and include 
the next generation of direct-search dark matter experiments and the 
Fermilab-led international long-baseline neutrino program. Planning to 
understand the infrastructure developments necessary to accommodate these 
future projects is well advanced and in some cases have already started.  
SURF is a dedicated research facility with significant expansion 
capability.
\end{abstract}


\section{Introduction}

Many disciplines benefit from access to an underground facility dedicated 
to scientific research, including physics, biology, geology, engineering, 
and a well-established science program is currently underway at the 
Sanford Underground Research Facility (SURF).  The unique underground 
environment at SURF allows researchers to explore a host of important 
questions regarding the origin of life and its diversity, mechanisms 
associated with earthquakes and a number of engineering topics.  A deep 
underground laboratory is also where some of the most fundamental topics 
in physics can be investigated, including the nature of dark matter, the 
properties of neutrinos and the synthesis of atomic elements within stars.

SURF is being developed in the former Homestake gold mine, in Lead, South 
Dakota~\cite{SURF-Heise-2014, SURF-Lesko, SURF-Heise-2010}. Barrick Gold 
Corporation donated the site to the State of South Dakota in 2006, 
following over 125 years of mining~\cite{Mitchell-2009}, during which time 
over 600 km of tunnels and shafts were created in the facility, extending 
from the surface to over 2450 meters (8000 feet) below ground.  The 
Laboratory property comprises 186 acres on the surface and 7700 acres 
underground, and the Surface Campus includes approximately 23,500 gross 
square meters (253,000 square feet) of existing structures. The South 
Dakota Science and Technology Authority (SDSTA) operates and maintains the 
Sanford Laboratory at the Homestake site in Lead, South Dakota with 
management and oversight by Lawrence Berkeley National Laboratory (LBNL).

\begin{figure}
\begin{center}
\includegraphics*[scale=0.36]{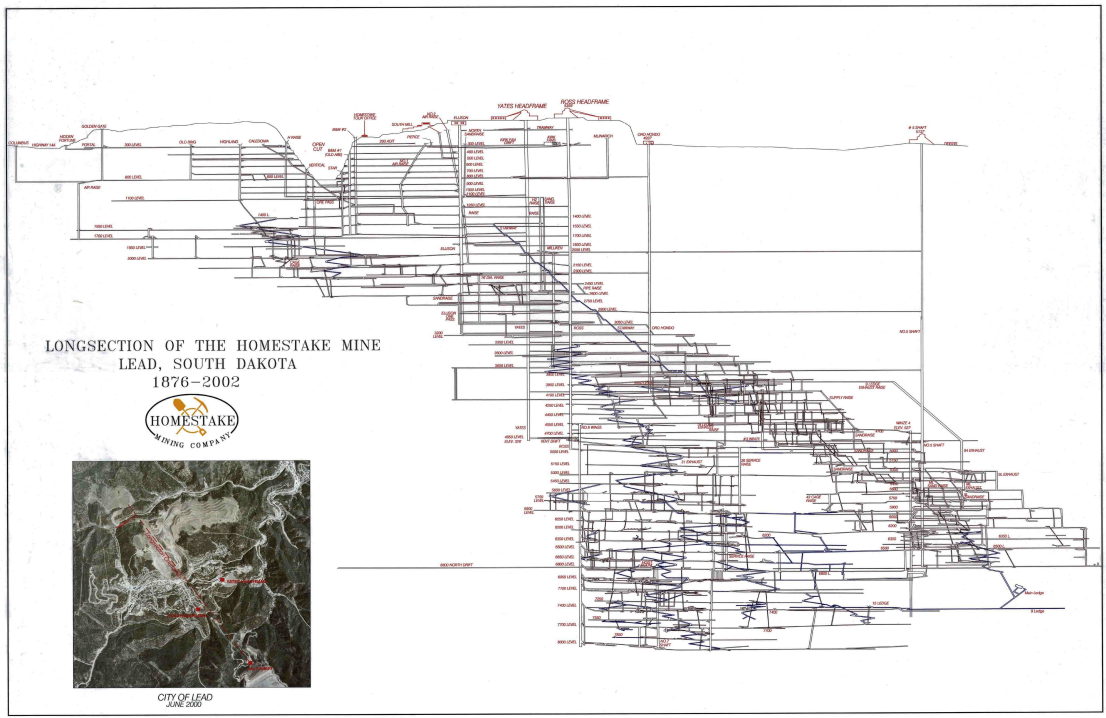}
\caption{\label{fig:Longsection} The long section of the former Homestake 
Gold Mine, in which the dark lines represent vertical shafts and 
horizontal tunnels projected onto a NW to SE plane.  This figure 
illustrates the 60 underground levels extending to greater than 2450 
meters (8000 feet) below ground. For scale, the horizontal length of the 
projection is 5.2~km.  The location of cross section is indicated in the 
inset.}
\end{center}
\end{figure}

In 2006, South Dakota philanthropist, T.\ Denny Sanford, gifted \$70M to 
convert the former mine into a research laboratory and develop a science 
education facility. With these funds and a strong commitment from the State 
of South Dakota (appropriations of \$42.2M to date), safe access to the 
underground has been reestablished and experimental facilities have been
commissioned and certified for occupancy.

The initial concepts for SURF were developed with the support of the 
U.S.\ National Science Foundation (NSF) as the primary site for the NSF's 
Deep Underground Science and Engineering Laboratory 
(DUSEL)~\cite{DUSEL-Lesko}. With the National Science Board's decision to 
halt development of a NSF-funded underground laboratory, the U.S.\ 
Department of Energy (DOE) now supports the majority of the operation of 
the facility. Support for experiments at SURF comes from both the NSF and 
DOE as well as other agencies such as the USGS and NASA.  Elements of the 
Homestake DUSEL Preliminary Design Report~\cite{DUSEL_PDR-Lesko} continue 
to be useful as the feasibility for portions of the original plan are 
investigated.


\section{Facility Operations Infrastructure}

Maintenance and operation of key elements of facility infrastructure 
enables safe access underground.  Transportation of personnel and 
materials underground is accomplished using the two primary shafts, the 
Ross Shaft and the Yates Shaft.  Pairs of hoists near both the Ross and 
Yates shafts move personnel and rock conveyances through the respective 
shafts.  Pumping stations in the Ross Shaft allow ground water to be 
pumped to surface.  Underground ventilation is provided by the Oro Hondo 
fan as well as the fan at \#5 Shaft, which bring fresh air underground via 
the Ross and Yates Shafts.

A key feature of the Sanford Laboratory is the capacity for redundancy.  
Redundant power, optical fiber and ventilation air are brought underground 
via both the Ross and Yates Shafts.  Multiple emergency egress options are 
provided by Ross and Yates shaft (and separate compartments therein) as 
well as by ramp systems that connect numerous underground levels.

Initial rehabilitation of the surface and underground infrastructure 
focused on the Ross shaft that provides access to the majority of 
underground utilities, including the pumping system used to remove water 
from the mine.  Contracted work in the Ross shaft started July 2008 and 
was completed by October 2008, after which SDSTA personnel continued with 
maintenance and renovations, including the removal of unused legacy piping 
(12 km) and power/communication cables (2 km).  Started in November 2008, 
the Yates shaft initial renovation was completed in May 2012 with the 
installation of a new personnel conveyance and emergency braking system.  
The Ross Shaft provided primary underground access from the start of 
re-entry in 2007 until the Yates shaft was ready in May 2012.  Personnel 
and materials are transported underground mainly using the Yates cage, 
which has dimensions 1.4~m wide $\times$ 2.7~m tall $\times$ 3.8~m long 
and has a maximum load capacity of 4540~kg (10,000~lbs); certain loads 
with widths of up to 1.7~m can be transported beneath the cage.  The Yates 
cage schedule accommodates three shifts per day for science personnel, 
providing 24-hour access as needed.  The maximum underground occupancy at 
SURF is determined by the number of personnel that can access a safe 
location in one hour.  Based on the current configuration of the various 
conveyances, the current underground occupancy limit is 72 persons.

In order to provide increased capacity to support the construction and 
operation of future experiments, state and private funds have been 
allocated to perform extensive renovations in the Ross Shaft.  
Refurbishment of the Ross Shaft began in August 2012, and over 670 meters 
(2200 feet) of new steel and associated ground support has been installed 
as of the end of December 2014.  A total of 1572~m (5159~ft) of new 
steel sets will be installed as part of the Ross shaft upgrade project, 
which is expected to be completed by mid-2017.  Once the Ross shaft is 
completed, a renovation of the Yates Shaft will be scheduled.


The average ground water inflow into the underground workings is 
approximately 730~gpm.  After pumping ceased in June 
2003~\cite{Murdoch-2012, Zhan_Duex-2010}, the mine filled with water until 
a high-water mark of 1381 meters (4530 feet) below surface was reached in 
August 2008.  Sustained pumping resumed in June 2008, dropping the water 
level below the 4850-foot level (4850L) by May 2009, after being flooded 
for an estimated 16 months.  Since April 2012 the water level is being 
maintained around the 1830-meter (6000-foot) level below surface.  While 
the potential for accommodating deeper access exists, without a funded 
mandate to develop laboratory space below the 4850L, there is benefit in 
terms of cost and safety for maintaining the water level around the 
current level.  If pumping now stopped, it would take an estimated 12--18 
months to impact infrastructure on the 4850L.

A deep-well pump was installed July 2010 and is currently located about 
1958 meters (6424 feet) below surface in \#6 Winze, which extends from the 
4850L to the deepest areas of the facility.  Permanent stations employing 
700-horsepower pumps are located on the 1250, 2450, 3650 and 5000 Levels 
in the Ross shaft.  Water received at the surface Waste Water Treatment 
Plant (WWTP) is combined with Homestake-Barrick water and treated to 
remove iron that has leached from the mine workings and trace amounts of 
ammonia.  The discharge capacity from the WWTP is roughly 2000 gpm using 
biological and sand-filter technologies.

Single-mode fiber optics cable is deployed throughout the facility and 
current network hardware provides inter-campus communication at 100--1000 
Mbps.  Redundant connections exist to the outside world, including 
commodity internet (``Internet 1'') at 1~Gbps and research internet 
(``Internet 2'' via the state Research, Education and Economic Development 
(REED) network) at 1~Gbps, which can be expanded to 100~Gbps with 
appropriate hardware upgrades.  Redundant fiber pathways connect the 
surface to the 4850L via the Ross and Yates Shafts, and work in underway 
to ensure that the core network equipment is protected by uninterruptible 
power supplies and in most cases generator power.  Researchers are making 
good use of the network infrastructure as illustrated in 
Table~\ref{tab:data}.  In 2014, the average daily throughput was 
approximately 710~GB.

\begin{table}[htbp]
\caption{\label{tab:data}Network traffic value increasing over time, 
indicating the evolution of science activities at SURF.  The majority of 
the data transfers to date have been associated with the LUX experiment. 
``Internet 2'' was used to route approximately 77\% of the data.}
\begin{center}
\begin{tabular}{lc}
\br
{\bf Year} & {\bf Total Data Traffic}  \\   
           & {\bf (TB)} \\
\mr
2009       & 2.6 \\ 
2010       & 5.4 \\
2011       & 12 \\
2012       & 22 \\
2013       & 249 \\
2014       & 259 \\
\br
\end{tabular}
\end{center}
\end{table}

\section{Surface Science Facilities}

\begin{figure}
\begin{center}
\includegraphics*[scale=0.60]{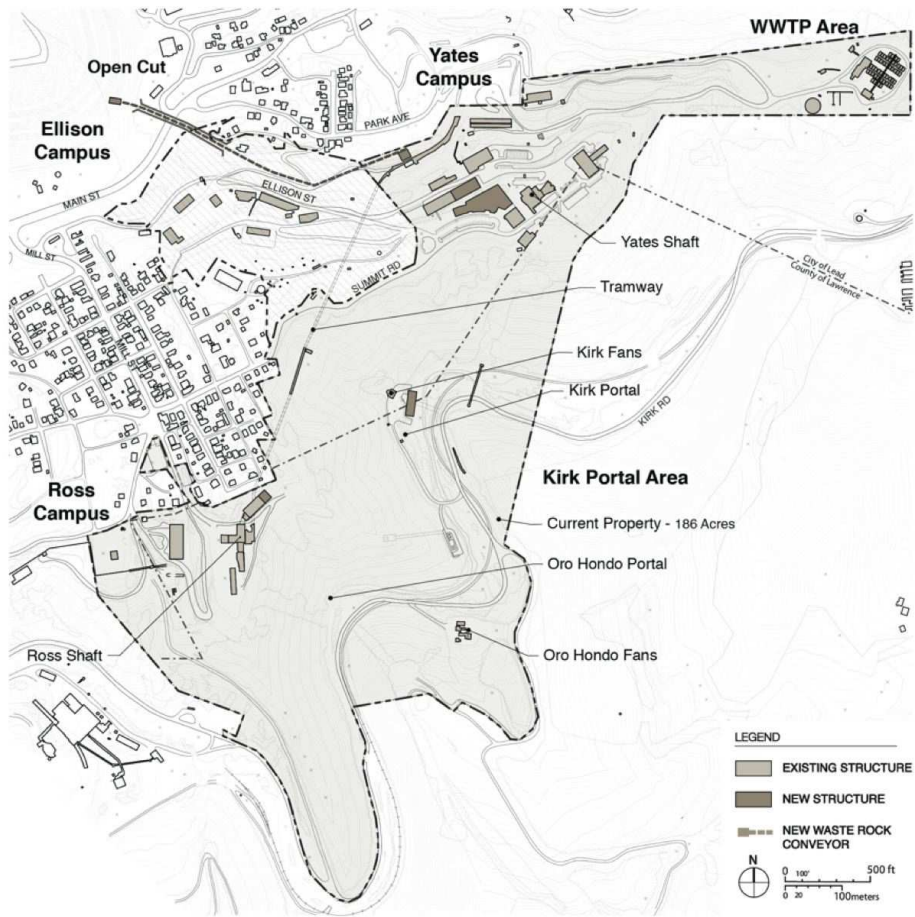}
\caption{\label{fig:SurfaceCampus} A plan view of the surface campus at 
SURF, which comprises 186 acres and includes approximately 23,500~m$^{2}$ 
(253,000~ft$^{2}$) of existing structures.  Locations of the Ross and 
Yates shafts are indicated as is the main ventilation fan (Oro Hondo) and 
the Waste Water Treatment Plant (WWTP).}
\end{center}
\end{figure}

In addition to the considerable underground extent at SURF, surface 
facilities exist at both the Yates and Ross surface campuses to facilitate 
science activities, including administrative support and office space, 
communication systems for education and public outreach, the WWTP for 
handling and processing waste materials and a warehouse for 
shipping/receiving.  A number of spaces are currently being used for 
storage as well as experiment preparation and construction activities.  
Figure~\ref{fig:SurfaceCampus} shows the extent of the surface property.

While there are many surface amenities, three main facilities directly 
serve science needs: the Core Archive, the Sawmill and the
Surface Laboratory, all of which are located at the Yates surface campus.  
Other surface buildings could be modified or renovated to meet specific 
needs.

\subsection{Core Archive}

Donated by Homestake-Barrick, SURF is the steward of 39,760 boxes of drill 
core rock\footnote{Diamond drilling produces cylinders of rock of various 
diameters called drill core.}, which corresponds to 2688 drill holes with 
a total length of approximately 91 km.  Homestake core holes extend to 
3290 meters (10,800 feet) below surface.  An additional 1646~m of core 
were added to the collection in Fall 2009 as part of the geotechnical 
investigations on the 4850L for DUSEL, and a further 770~m of core have 
also been collected for geotechnical investigations for the LBNF project 
in April 2014.

The SD Geological Survey has assisted with the development of an online 
database that so far includes 58,000+ entries, representing 1740 drill 
holes.  

\subsection{Sawmill}

A former sawmill is currently being used by SURF personnel as well as 
researcher groups for various tasks.  For instance, the {\sc Majorana} 
collaboration is performing basic preparation, staging and construction 
activities in that facility.

\subsection{Surface Laboratory}

\begin{figure}
\begin{center}
\includegraphics*[scale=0.16]{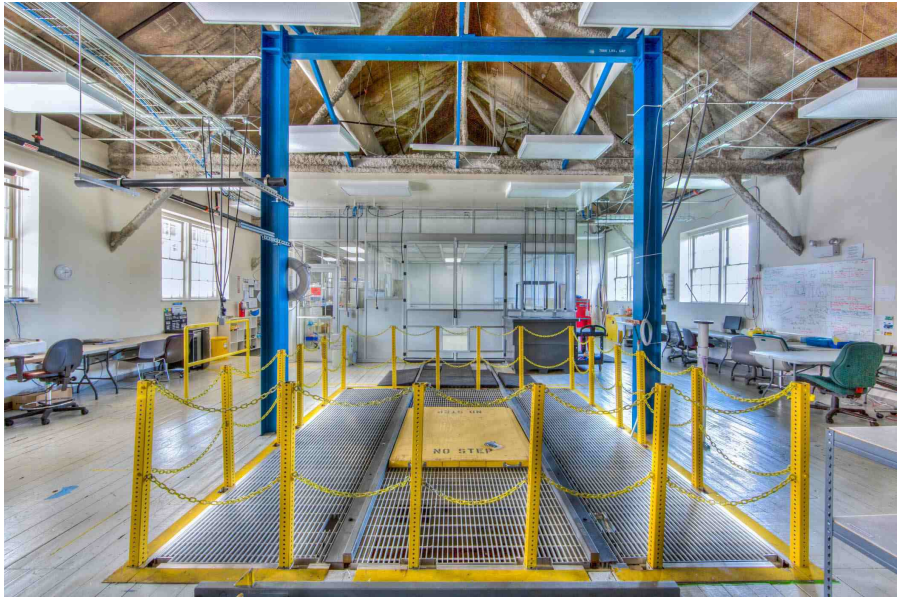}
\caption{\label{fig:SurfaceLab} Surface Laboratory. The recessed area 
with the tank is under the grating in the foreground and the cleanroom is 
located in the background.  A hoist can be installed on the blue I-beam 
structure.}
\end{center}
\end{figure}

Renovations were undertaken in 2009 in order to transform a former 
warehouse into a laboratory.  Construction was completed in early 2010, 
resulting in approximately 190 m$^{2}$ of lab space in the top-most level 
of a four-story building as shown in Figure~\ref{fig:SurfaceLab}.  The 
facility includes a cleanroom (5.6~m $\times$ 6.6~m with a 2.7~m ceiling 
height) and corresponding dedicated air handling and filtration system as 
well as a tank that can be used as a water shield (2.8~m diameter $\times$ 
4~m high).  The tank is installed in a recessed shaft in the center of the 
laboratory space.  The laboratory was initially designed to meet the needs 
of the LUX experiment, but is now used to support multiple research 
groups.  Additional renovations to accommodate the LUX-ZEPLIN (LZ) 
experiment are expected to begin as early as Summer 2015.

\section{Underground Science Facilities}

A number of levels required to support facility operations infrastructure 
can also accommodate research activities. However, the main infrastructure 
for the support of science activities has been developed on the 4850L with 
formal campuses located near both the Ross and Yates shafts.

\subsection{4850L Ross Campus}

\begin{figure}
\begin{center}
\includegraphics*[scale=0.58]{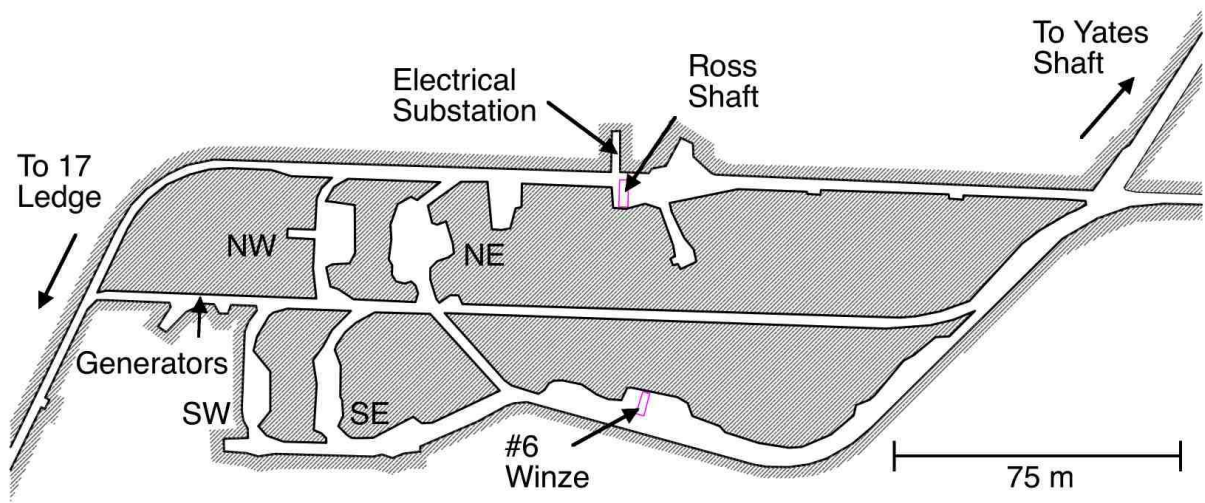}
\caption{\label{fig:RossCampus} 4850L Ross Campus.  Four existing 
excavations are labeled: NW, NE, SE, SW. The two western shop areas are 
currently in use, whereas the two eastern shops are presently undergoing 
renovations.}
\end{center}
\end{figure}

The 4850L Ross Campus includes the Ross Shaft and \#6 Winze and 
encompasses a set of four existing excavations that were used as 
maintenance shops during mining activities.  These former shops afford an 
economical means to implement experiments or other equipment in a timely 
manner.  The layout is shown in Figure~\ref{fig:RossCampus}, including the 
location of the current electrical substation\footnote{The capacity of the 
current electrical substation near the Ross shaft is sufficient for the 
planned new developments at the Ross Campus.  LBNF will require a separate 
dedicated substation.} and generators.  The two western shop areas are 
currently in use, whereas the two eastern shops are presently undergoing 
renovations.

\begin{table}[htbp]
\caption{\label{tab:RossCampusVol}Footprint areas and volumes for the 
4850L Ross Campus spaces.  The volume for the NE area is derived from 
laser-scan data.}
\begin{center}
\begin{tabular}{lcc}
\br
{\bf Ross Campus Location} & {\bf Area}      & {\bf Volume}  \\
                           & {\bf (m$^{2}$)} & {\bf (m$^{3}$)} \\
\mr
NW ({\sc Majorana} Electroforming)  & 184   &  504 \\ 
NE (BHUC)                           & 297   &  707 \\ 
SE (CASPAR)                         & 137   &  376 \\ 
SW (Refuge Chamber)                 & 121   &  332 \\ 
\mr
{\bf Total}                    & {\bf 739}  & {\bf 1919} \\
\br
\end{tabular}
\end{center}
\end{table} 

\begin{figure}
\begin{center}
\includegraphics*[scale=0.5]{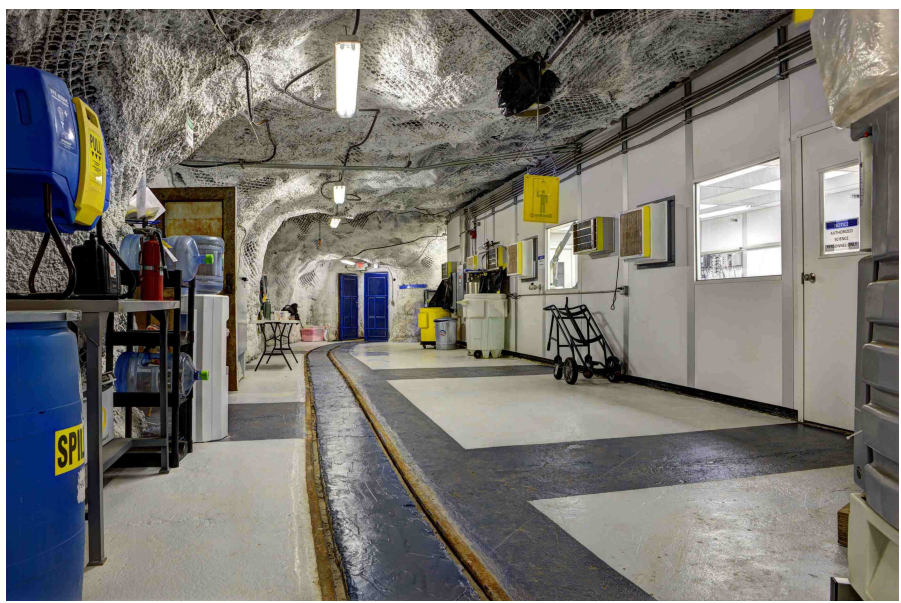}
\caption{\label{fig:RossCampus-MJD} The {\sc Majorana} electroforming 
laboratory at the 4850L Ross Campus.  The cleanroom is located on the 
right-hand side.}
\end{center}
\end{figure}

The NW shop area was renovated starting in December 2009 so that a 
cleanroom (3.7~m wide $\times$ 12.2~m long, with an overall height of 
3.1~m and an interior ceiling height of 2.5~m) for producing ultra-pure 
copper could be installed to support the {\sc Majorana} experiment 
schedule.  After several phases of development, the laboratory area, 
complete with fire suppression system, was available for occupancy in 
March 2011 as shown in Figure~\ref{fig:RossCampus-MJD}.  Production copper 
electroforming operations for the {\sc Majorana Demonstrator} began in 
July 2011 and are expected to continue through Spring 2015.


\begin{figure}
\begin{center}
\includegraphics*[scale=0.28]{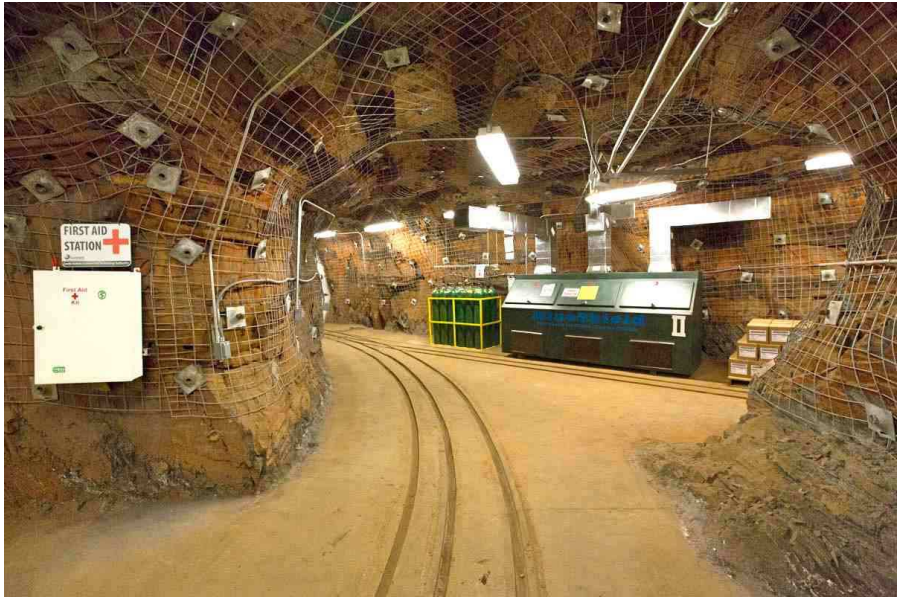}
\caption{\label{fig:RossCampus-RefugeChamber} The Refuge Chamber at the 
4850L Ross Campus can accommodate 72 people for 96 hours. Amenities 
shown include first-aid kits, one of two CO$_{2}$ scrubbing units and a 
bank of compressed air cylinders as well as water and rations.} 
\end{center}
\end{figure}

Coincident with the start of the Ross Shaft rehabilitation in the fall of 
2012, the SW area was converted into a safety Refuge Chamber that can 
accommodate 72 people (current maximum occupancy) for up to 96 hours.  It 
includes air locks at the two entrances, compressed air, CO$_{2}$ 
scrubbers, air conditioners, rations and communications.  Some of the 
amenities are shown in Figure~\ref{fig:RossCampus-RefugeChamber}.

\begin{figure}
\begin{center}
\includegraphics*[scale=0.30]{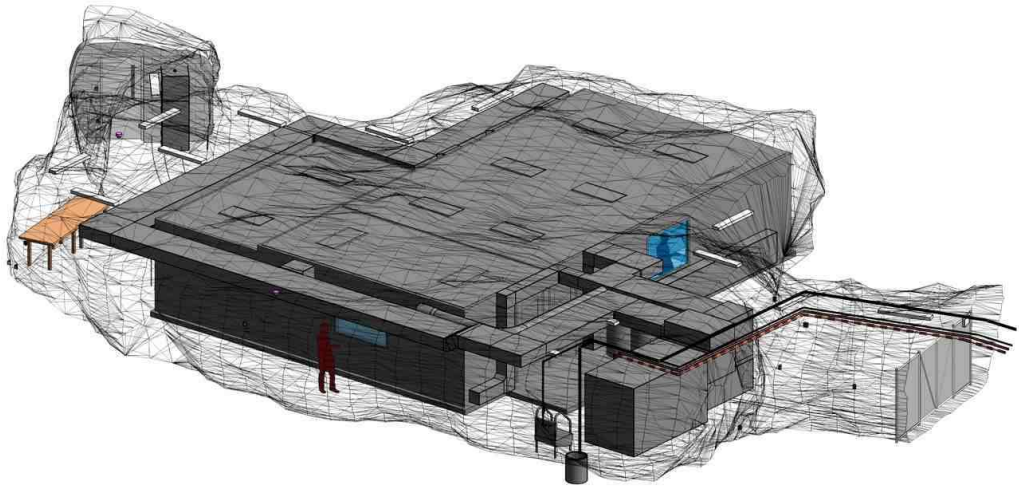}
\caption{\label{fig:RossCampus-BHUC} The Black Hills State University 
Underground Campus at the 4850L Ross Campus. Beneficial occupancy is 
expected in Summer 2015.}
\end{center}
\end{figure}

The NE shop will host the Black Hills State University Underground Campus 
(BHUC) that is intended to support multidisciplinary research from 
multiple institutions~\cite{BHUC}.  Installation of additional ground 
support was completed in December and shotcrete will be applied starting 
in January, with beneficial occupancy anticipated for Summer 2015.  The 
main feature of the campus will consist of a 268-m$^2$ cleanroom as 
illustrated in Figure~\ref{fig:RossCampus-BHUC}.  


The Compact Accelerator System for Performing Astrophysical Research 
(CASPAR) will be installed in a laboratory area comprising 401~m$^2$, 
which includes the SE shop area as well as part of an adjoining tunnel.  
An existing accelerator at the University of Notre Dame will be 
transported to SURF and installed underground starting in Summer 2015.


\begin{figure}
\begin{center}
\includegraphics*[scale=0.30]{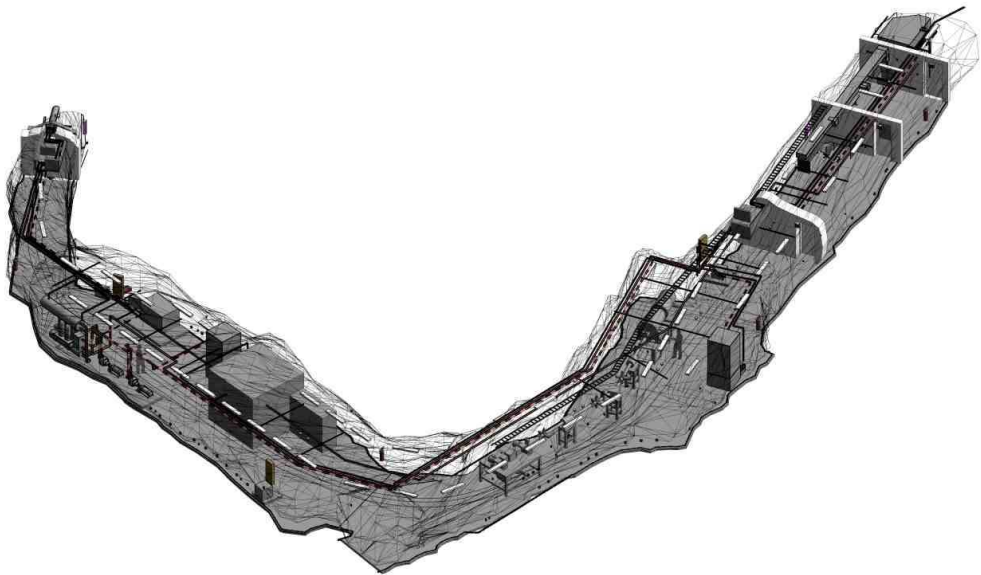}
\caption{\label{fig:RossCampus-CASPAR} The Compact Accelerator System for 
Performing Astrophysical Research (CASPAR) will be installed in the SE 
shop area of the 4850L Ross Campus. An existing accelerator at the 
University of Notre Dame will be transported to SURF and installed 
underground starting in Summer 2015.}
\end{center}
\end{figure}

\subsection{4850L Davis Campus}

\begin{figure}[h]
\begin{minipage}{18pc}
\includegraphics*[scale=0.216]{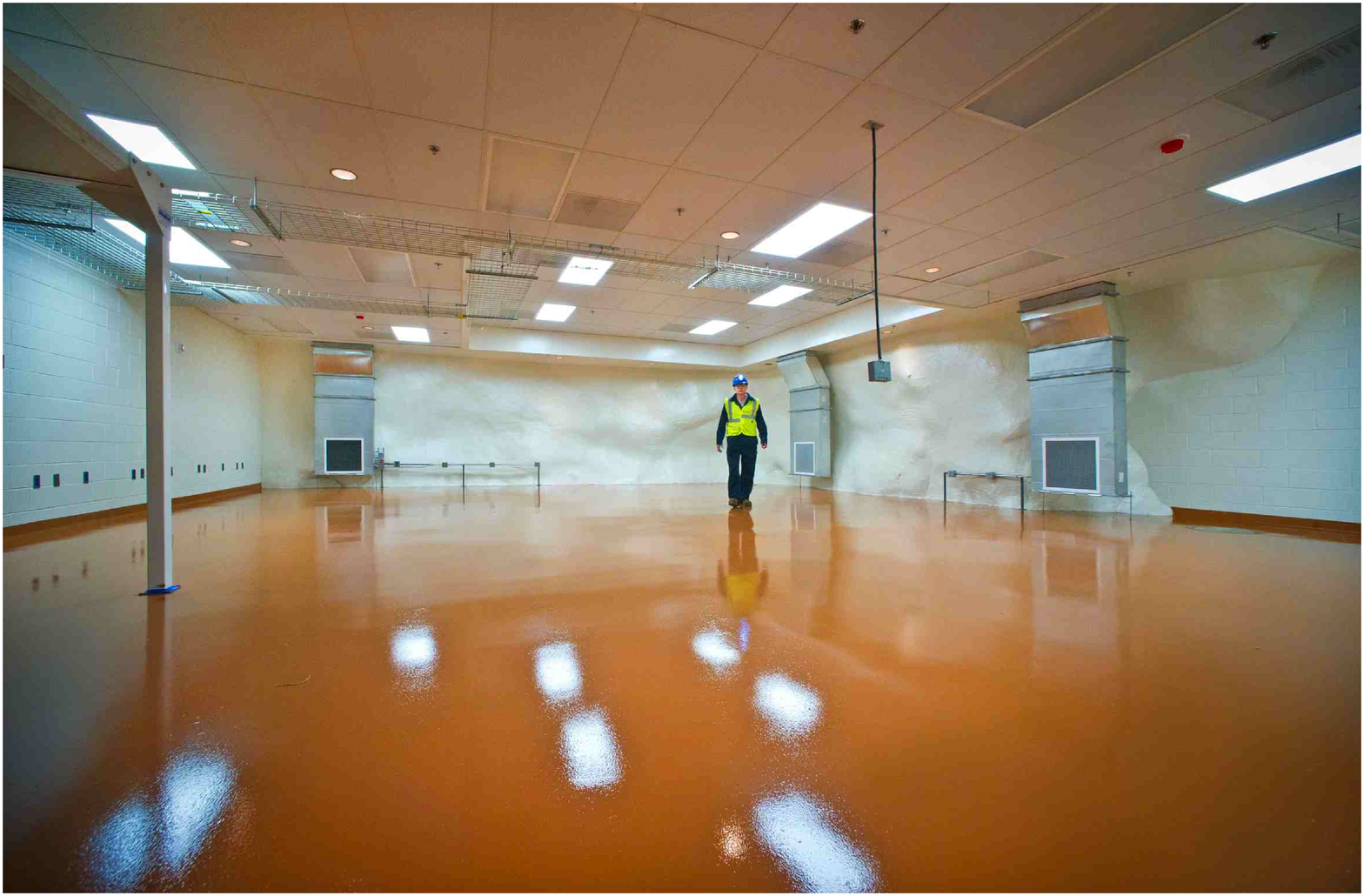}
\end{minipage}\hspace{2.5pc}%
\begin{minipage}{18pc}
\includegraphics*[scale=0.20]{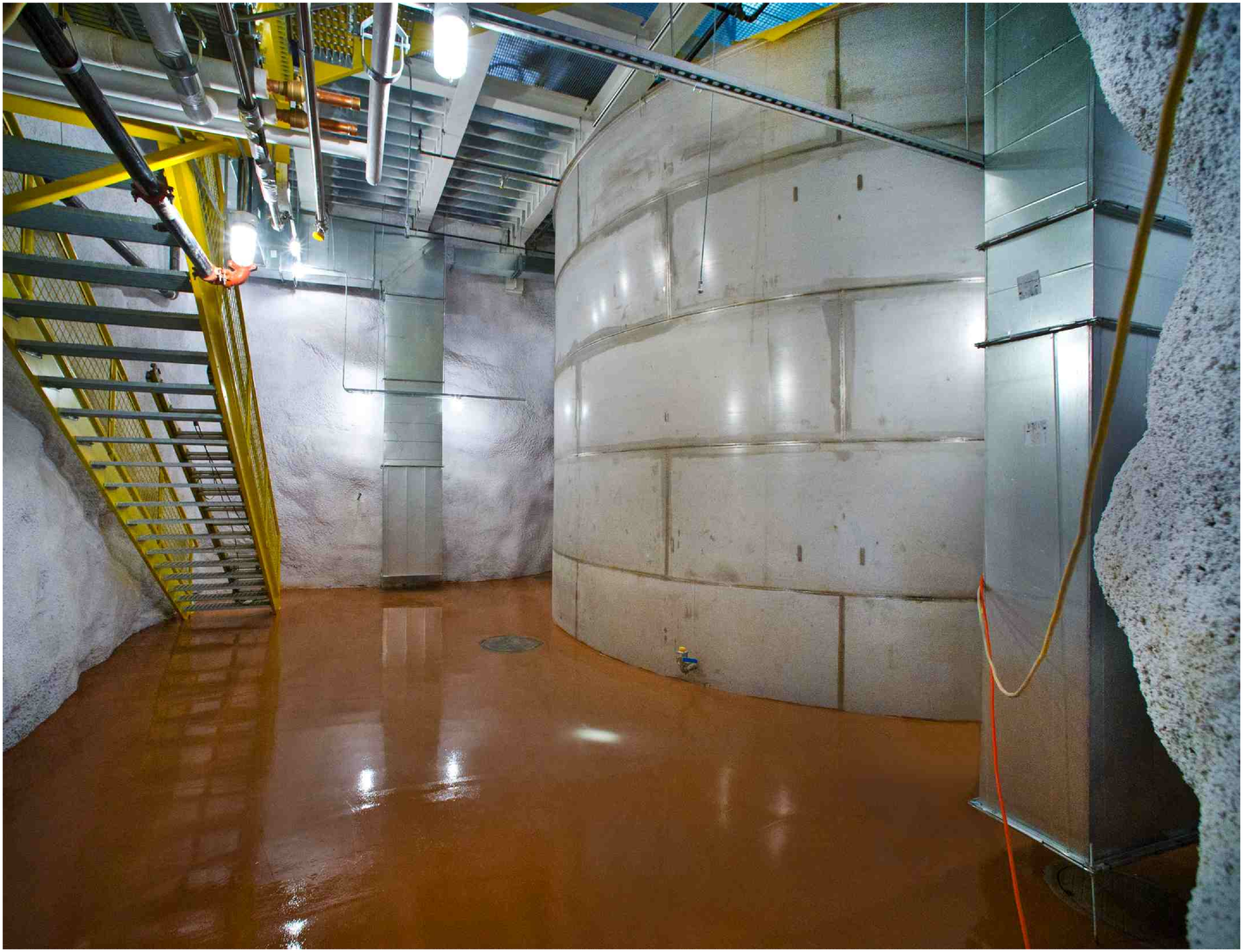}
\end{minipage}
\caption{\label{fig:DavisCampusPics}Pictures of the Davis Campus in May 
2012 after outfitting had finished: (left) {\sc Majorana Demonstrator} 
Detector Room, (right) water shielding tank in Lower Davis for the LUX 
detector.  A considerable amount of equipment has since been installed by 
both collaborations.}
\end{figure}

A state-of-the-art laboratory complex called the Davis Campus has been 
constructed at the 4850L near the Yates Shaft.  The Davis Campus 
represents a \$16M South Dakota commitment using state and private funds.  
New excavation for the Davis Campus took place during September 2009 -- 
January 2011, during which time 16,632 tonnes (18,334 tons) of rock was 
removed.  Rather than being transported to the surface, areas were 
identified on the 5000L (via the 4850L) for rock storage.  Shotcrete was 
applied in both the Davis Cavern (average thickness 12.7~cm) and the 
Transition space (average thickness 8.9~cm).  Laboratory outfitting began 
in June 2011 and was substantially complete in May 2012; pictures are 
shown in Figure~\ref{fig:DavisCampusPics}.  A final occupancy certificate 
was issued in July 2012, which specifies the maximum number of occupants 
to be 62 persons.

\begin{figure}
\begin{center}
\includegraphics*[scale=0.58]{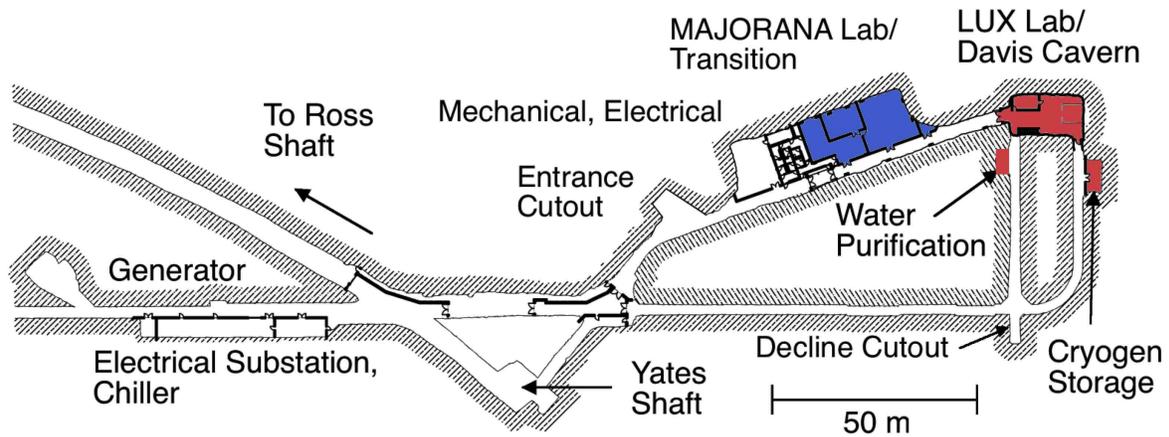}
\caption{\label{fig:DavisCampus} 4850L Davis Campus showing the main 
laboratory spaces and the proximity to the Yates Shaft, which is 
approximately 180~m from the main laboratory entrance.  {\sc Majorana} 
spaces are shown in blue whereas the LUX areas are red.  The Davis Cavern 
offers two floors of laboratory space for LUX.}
\end{center}
\end{figure}

The two main experiments areas are the {\sc Majorana} Lab/Transition space 
(43~m L $\times$ 16~m W $\times$ 4~m H) and the Davis Cavern (17~m L 
$\times$ 10~m W $\times$ 12~m H), as shown in 
Figure~\ref{fig:DavisCampus}.  Quoted dimensions are average values based 
on post-shotcrete laser-scan data.  The Davis Campus footprint consists of 
3017 m$^{2}$ (32,478 ft$^{2}$) total space as indicated in 
Table~\ref{tab:DavisCampusVol}.  While space is at a premium at the Davis 
Campus, two cutouts exist outside the clean space offering footprints of 
30--33 m$^{2}$, with an average ceiling height of 3.7~m.  These spaces 
could be made available to groups with modest equipment needs.


\begin{table}[htbp]
\caption{\label{tab:DavisCampusVol}Footprint areas and volumes for the
4850L Davis Campus spaces.  The volumes are derived from post-shotcrete 
(where appropriate) laser-scan data with finished flooring.}
\begin{center}
\begin{tabular}{lcc}
\br
{\bf Davis Campus Location}  & {\bf Area}      & {\bf Volume}  \\
                             & {\bf (m$^{2}$)} & {\bf (m$^{3}$)} \\
\mr
LUX                          &  375        &  1976 \\ 
{\sc Majorana Demonstrator}  &  300        &  1279 \\  
Low-Bkgd Counting            &   33        &   140 \\
Common Science               &  244        &   980 \\
Other/R\&D Science           &   63        &   253 \\
Facility Infrastructure      &  508        &  2063 \\
Access Drifts                & 1494        &  4664 \\
\mr
{\bf Total}                  & {\bf 3017}  & {\bf 11,354} \\
\br
\end{tabular}
\end{center}
\end{table}

\subsubsection{Davis Campus Monitoring}

Monitoring is conducted for a variety of gases at the Davis Campus.  Due 
to the use of cryogens in various sections of the laboratory, a total of 
nine oxygen sensors are installed throughout the Davis Campus.  
Furthermore, a half-dozen portable gas testers sensitive to CO, H$_{2}$S, 
O$_{2}$ and CH$_{4}$ are also available throughout the campus.  A fixed 
carbon monoxide monitor is located near the Yates Shaft, by the air intake 
for the Davis Campus.  Separate heat and smoke sensors are installed in 
all areas of the Davis Campus.

A general facility alarm is activated throughout the Davis Campus in the 
event that any of the following occur: low oxygen, heat or smoke detected, 
fire suppression water is flowing or a manual pull station is activated.  
Currently, regular evacuations exercises are held biweekly, with more 
comprehensive scenarios conducted quarterly.

An array of extensometers is installed to monitor ground motion and 
convergence at the Davis Campus -- data collected so far indicate nothing 
unexpected.  Geotechnical interpretations were performed using drill core 
and other rock mechanics studies~\cite{Pariseau-2012}.

\subsubsection{Davis Campus Services}

Services provided throughout the Davis Campus include fire sprinklers, 
potable and non-potable (industrial) water, lighting (including emergency 
lighting), ventilation and air conditioning. A building management system 
provides controls throughout the Campus. Cooling is provided with two 
redundant 50-ton (633-MJ) chillers supplying chilled water to three air 
handling units that provide ventilation to separate campus spaces. Chilled 
water and a facility air compressor are also available for experiments to 
connect equipment directly. A dedicated 1500~kVA substation provides 
sufficient capacity for the experiment and facility needs, with margin for 
future expansion (e.g., LZ and any scaling of the {\sc Majorana 
Demonstrator}). Emergency power for lighting is provided with batteries in 
the lighting system to provide immediate light, while a standby diesel 
generator near the campus provides up to 24 hours of power to all safety 
systems in the campus. This includes water pumps in the nearby Yates shaft 
to prevent water from rising into the campus spaces.  Three separate air 
handling units provide multi-stage air filtration for the Davis Campus 
clean spaces and establish relative pressure balances to help maintain 
appropriate relative levels of cleanliness.  Starting in July 2014, the 
intake air for the Davis Campus clean area air handling system was 
pre-processed using a dedicated dehumification system.

\subsubsection{Davis Campus Cleanliness}

Cleanliness at the Davis Campus is maintained through a combination of 
transition zones and protocols as well as dedicated cleaning staff.  When 
supplies and equipment are brought from the dirty space into the clean 
space, items either need to be suitably enclosed or cleaned in the 
facility entrance Cart Wash.  Personnel typically enter though a series of 
rooms where outer coveralls and dirty-side gear are removed in favor of 
corresponding clean items.  Laundry services are also available in the 
clean space.  The facility design includes two sets of personnel showers, 
but so far they have not been necessary.

When the laboratory is occupied, particle counts ($>$0.5 $\mu$m per cubic 
foot) in all main common areas are below 10,000/ft$^{3}$.  The {\sc 
Majorana} collaboration follows additional gowning protocols to achieve 
ultra-low cleanliness levels in their laboratory spaces.  Particle counts 
have been monitored at the Davis Campus since September 2012, and a 
summary of cleanliness states throughout various areas is presented in 
Table~\ref{tab:cleanliness}.

Dust from the air-handling unit filters as well as floor sweepings have 
been collected and routinely analyzed for more than 6 months.  The main 
activity contributions appear to be from $^{210}$Pb and $^{7}$Be.  For 
example, {\it in situ} dust has the following average levels: $\sim$1~ppm 
U, $\sim$2 ppm Th and 1--2\% K~\cite{BLBF_report}.

\begin{table}[htbp]
\caption{\label{tab:cleanliness}Median particle count levels in various 
Davis Campus laboratory locations during occupied periods.  Additional 
cleanliness protocols are instituted in the MJD areas.}
\begin{center}
\begin{tabular}{lcc}
\br
{\bf Davis Campus Location}       & {\bf Particle Count} \\
                                  & {\bf (0.5~$\mu$m/ft$^{3}$)} \\
\mr
Entry/Lockers (MJD)               & 790 \\
Common Transition Hallway         & 2320 \\
Common Corridor                   & 3240 \\
Davis Cavern, Lower (LUX)         & 3400 \\
Davis Cavern, Lower (CUBED)       & 720 \\
Davis Cavern, Lower (BLBF)        & 490 \\
Davis Cavern, Upper (LUX)         & 400  \\
Detector Room (MJD)               & 100--400 \\
\br
\end{tabular}
\end{center}
\end{table}

\section{Facility Characteristics}

A number of specific properties that define and influence capabilities for 
various underground laboratory spaces are reviewed in the sections below, 
including the geographic coordinates, geology and rock overburden 
characteristics as well as radon and radioactivity considerations.

\subsection{Coordinates}

Initial survey work in 2010 projected the Universal Transverse Mercator 
(UTM) coordination system (Zone 13 North, NAD83 datum with 2002 Epoch) 
underground to benchmarks on the 4850L.  Projected Easting (x) and 
Northing (y) coordinates for various underground locations are presented 
in Table~\ref{tab:coord}.  Future surveys in support of the long baseline 
neutrino initiative will improve on these preliminary efforts.

\begin{table}[htbp]
\caption{\label{tab:coord}Coordinates estimates for various locations of 
interest.  Both UTM and geographical coordinates are presented.  U.S.\ 
survey feet are converted to meters.}
\begin{center}
\begin{tabular}{lcccc}
\br
{\bf Location}  & \multicolumn{2}{c}{\bf UTM Coordinates} &
\multicolumn{2}{c}{\bf Geographic Coordinates} \\
                & {\bf Easting} & {\bf Northing} &
{\bf Latitude} & {\bf Longitude} \\
                & {\bf (m)} & {\bf (m)} & & \\
\mr
\multicolumn{4}{l}{\bf 4850L Davis Campus} \\
\mr
LUX Detector         & 599527 & 4911873 & 44$^{\circ}$ 21' 11.88'' N & 
103$^{\circ}$ 45' 04.27'' W \\ 
MJD Detector         & 599516 & 4911838 & 44$^{\circ}$ 21' 10.75'' N & 
103$^{\circ}$ 45' 04.77'' W \\ 
Entrance Cutout      & 599531 & 4911781 & 44$^{\circ}$ 21' 08.92'' N & 
103$^{\circ}$ 45' 04.15'' W \\ 
\mr
\multicolumn{4}{l}{\bf 4850L Ross Campus} \\
\mr
MJD Electroforming   & 598939 & 4911087 & 44$^{\circ}$ 20' 46.70'' N & 
103$^{\circ}$ 45' 31.33'' W \\ 
BHUC                 & 598962 & 4911085 & 44$^{\circ}$ 20' 46.65'' N & 
103$^{\circ}$ 45' 30.31'' W \\ 
CASPAR               & 598939 & 4911045 & 44$^{\circ}$ 20' 45.36'' N & 
103$^{\circ}$ 45' 31.39'' W \\ 
\mr
\multicolumn{4}{l}{\bf 4850L Other} \\
\mr
LBNF (10 kt Cavern)  & 599236 & 4911041 &  44$^{\circ}$ 20' 45.07'' N & 
103$^{\circ}$ 45' 17.96'' W \\ 
LBNF (30 kt Cavern)  & 599223 & 4910960 & 44$^{\circ}$ 20' 42.44'' N & 
103$^{\circ}$ 45' 18.62'' W \\ 
Expt Hall (Generic)  & 599437 & 4911408 & 44$^{\circ}$ 20' 56.88'' N & 
103$^{\circ}$ 45' 08.66'' W \\ 
\mr
\multicolumn{4}{l}{\bf Other} \\
\mr
800L (Muon site~\cite{Bkgd-Muon})  & 598928 & 4911221 & 44$^{\circ}$ 20' 
51.07'' N & 103$^{\circ}$ 45' 31.74'' W \\ %
2000L (Muon site~\cite{Bkgd-Muon}) & 598764 & 4912928 & 44$^{\circ}$ 21' 
46.47'' N & 103$^{\circ}$ 45' 37.98'' W \\ %
\br
\end{tabular}
\end{center}
\end{table}

\subsection{Geology and Rock Overburden}

When considering the muon flux at key laboratory spaces on the 4850L, six 
geological formations are important to consider: Grizzly, Flagrock, 
Northwestern, Ellison, Homestake and Poorman.  The Yates Member (Unit) is 
the lowest stratigraphic unit of the Poorman Formation and is important 
for developments on the 4850L; other geological formations also have 
amphibolite manifestations that are accounted for in the analysis below.  
In addition, tertiary Rhyolite dikes occur throughout the rock mass.

\begin{figure}
\begin{center}
\includegraphics*[scale=0.35]{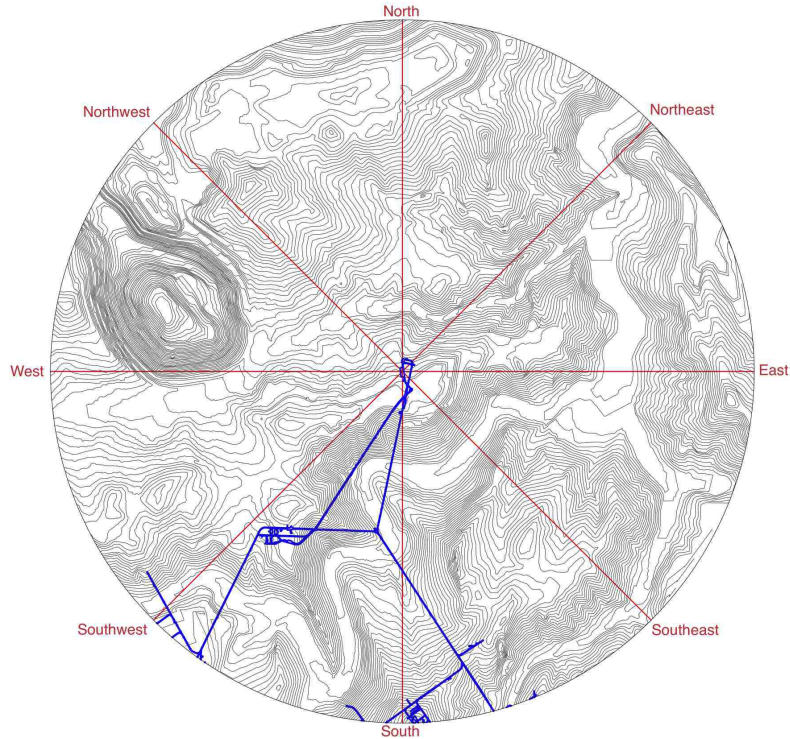}
\caption{\label{fig:Topo_4850L} Topological contours for the opening of a 
45-degree cone overlaid on top of an outline of the 4850L (shown in blue), 
centered on the Davis Campus.  Various cardinal coordinates are indicated 
with red lines.  The Open Cut is a prominent feature on the surface toward 
the west.  Contour lines indicate 6-meter (20-foot) elevation changes.}
\end{center}
\end{figure}

The surface topology varies significantly throughout the laboratory 
property as shown in Figure~\ref{fig:Topo_4850L}.  Using average rock 
densities and a 3-dimensional geological model~\cite{RoggenthenHart-2014} 
in addition to recent topological surveys, estimates of the rock 
overburden for various locations are summarized in Table~\ref{tab:mwe}. 
Errors in the vertical rock depth are expected to be modest ($\pm$10~m).  
To estimate the overburden uncertainties associated with the geological 
model, rock formation contact boundaries were adjusted for several 
locations based on the relative information about their locations in 
order to maximize and minimize the total density-scaled depths.  Density 
errors (1--6\%) were also factored in using $\pm$1~$\sigma$ (standard 
deviations), and contributed the majority of the uncertainty.  However, 
in addition to simply quantifying the measurement error, the 
uncertainties also characterize density variations due to differences in 
mineralogy in a given rock formation.  As a result of the combination, 
the average uncertainty in the water-equivalent overburden values was 
estimated to be $\sim$4\% for most of the locations; the overburden error 
for the 800L location is lower at roughly $\sim$2.5\%.  In addition to 
the vertical overburden depth values, the rock density averaged over a 
45-degree cone (determined by sampling the model at several discrete 
points through the cone volume) is presented for various underground 
locations as a way of characterizing the effective overburden over a 
broader extent.  Geochemical composition data for various formations are 
available from sources such as the USGS~\cite{Caddey-1991} and others.

\begin{figure}
\begin{center}
\includegraphics*[scale=0.285]{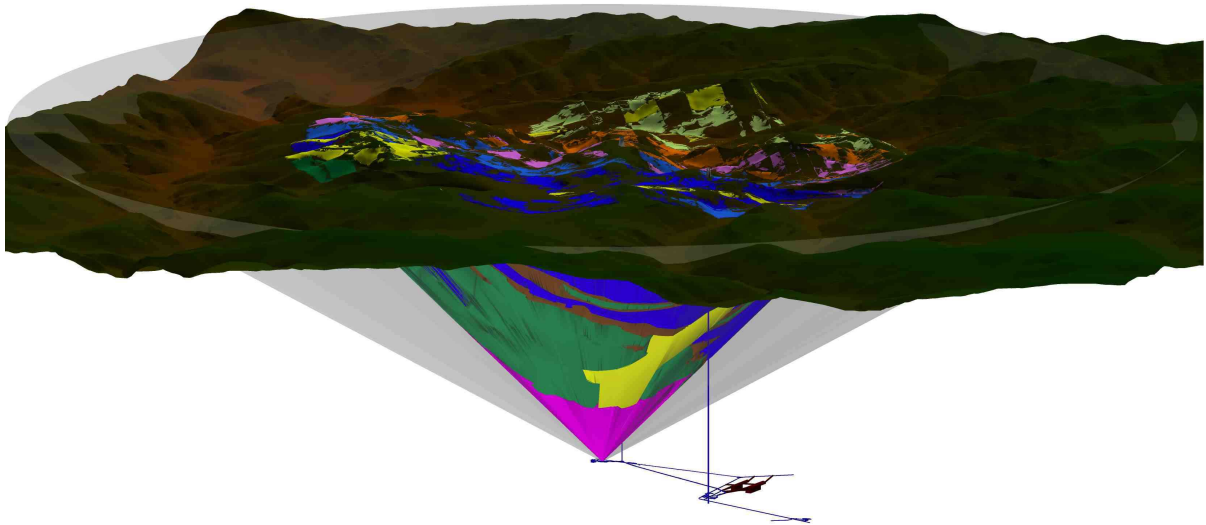} 
\caption{\label{fig:GeoCone}3D geological model showing rock cone above 
the 4850L centered on the Davis Campus as well as the surface topology.  
The colored section represents different rock formations in the 45-degree 
cone; the gray outline represents the extent of a 60-degree cone.  The 
vertical lines show the locations of the Ross and Yates shafts.}
\end{center}
\end{figure} 

\begin{table}[htbp] 
\caption{\label{tab:mwe}Rock overburden estimates for various underground 
locations at SURF, using a 3-dimensional geological 
model~\cite{RoggenthenHart-2014} and including depths scaled by density to 
give the meters of water equivalent (m.w.e.).  The average rock density is 
based on a 45-degree cone centered above the specific site.  The data 
reflect considerable variations in surface topology as well as geology.  
Properties related to the locations of recent muon flux measurements are 
also included. Uncertainties on the rock depth values are estimated to be 
$\pm$10~meters, whereas the estimated uncertainties in the density-scaled 
overburden values are $\sim$4\%; the overburden error for the 800L 
location is lower at roughly $\sim$2.5\%.}
\begin{center} 
\begin{tabular}{lccc} \br {\bf Location} & \multicolumn{2}{c}{\bf 
Vertical} & {\bf Average} \\
                & \multicolumn{2}{c}{\bf Rock Overburden} & {\bf Cone Density} \\
                & {\bf (m)} & {\bf (m.w.e.)} & {\bf (g/cm$^3$)} \\
\mr
\multicolumn{4}{l}{\bf 4850L Davis Campus} \\
\mr
LUX Detector         & 1466 & 4210  & 2.85 \\ 
MJD Detector         & 1477 & 4260  & 2.86 \\ 
Entrance Cutout      & 1494 & 4300  & 2.85 \\ 
\mr
\multicolumn{4}{l}{\bf 4850L Ross Campus} \\
\mr
MJD Electroforming   & 1503 & 4290  & 2.81 \\ 
BHUC                 & 1503 & 4380  & 2.83 \\ 
CASPAR               & 1499 & 4170  & 2.81 \\ 
\mr
\multicolumn{4}{l}{\bf 4850L Other} \\
\mr
LBNF (10 kt Cavern)  & 1384 & 3910  & 2.82 \\ 
LBNF (30 kt Cavern)  & 1375 & 3840  & 2.82 \\ 
Expt Hall (Generic)  & 1189 & 3390  & 2.80 \\ 
\mr
\multicolumn{4}{l}{\bf Other} \\
\mr
800L (Muon site~\cite{Bkgd-Muon})  & 283 & 770  & 2.73 \\ 
2000L (Muon site~\cite{Bkgd-Muon}) & 624 & 1700 & 2.74 \\ 
\br
\end{tabular}
\end{center}
\end{table}

\subsection{Radioactivity and Radon}

SURF strives to provide the lowest possible radioactivity environment for 
experiments hosted within the facility. This commitment had been 
integrated into the site preparation process from the early days of the 
facility design, and carried over to the realization of the 4850L Davis 
Campus laboratories.  These efforts include site and environmental 
characterization including rock radioactive measurements, use of 
low-radioactivity construction materials, and regular monitoring of 
environmental factors, including airborne radon.

The natural abundance of U/Th/K in Homestake rock is generally low, 
especially in certain geological formations such as the Yates 
Amphibolite, which have been measured to contain sub-ppm levels of U/Th.  
Samples from other formations such as the Rhyolite intrusions can be 
30--40$\times$ higher as illustrated in Table~\ref{tab:uthk}.  While not 
as low as the host Yates Amphibolite rock, the lowest-activity 
construction materials for the Davis Campus were selected from a large 
number of samples~\cite{Assay-Whitepaper, Assay-Oroville, Assay-YDC}.

\begin{table}[htbp]
\caption{\label{tab:uthk} Partial U/Th/K assay results for relevant SURF 
rock samples as well as key construction materials used at 
various laboratories~\cite{Assay-Whitepaper, Assay-Oroville, Assay-YDC}.}
\begin{center}
\begin{tabular}{lccc}
\br
              & {\bf Uranium} & {\bf Thorium}  & {\bf Potassium} \\ 
              & {\bf (ppm)}   & {\bf (ppm)}    & {\bf (\%)} \\
\mr
\multicolumn{4}{l}{\bf 4850L Davis Campus} \\
\mr
Yates Amphibolite Fm (Majority)  & 0.22  & 0.33  & 0.96 \\
Rhyolite Dike                    & 8.75  & 10.86 & 4.17 \\[0.2cm]
Shotcrete -- Low Activity        & 1.52  & 2.17  & 0.55 \\
Shotcrete -- Standard            & 2.00  & 3.35  & 1.23 \\
Shotcrete -- Finish Coat         & 1.62  & 3.08  & 0.79 \\
Masonry Blocks                   & 2.16  & 3.20  & 1.23 \\
\mr
\multicolumn{4}{l}{\bf 4850L Ross Campus} \\
\mr
Poorman Fm (Majority)            & 2.58  & 10.48 & 2.12 \\
Siderite/Grunerite Ironstone     & 0.41  & 2.30  & 0.25 \\
Shotcrete -- Standard            & \multicolumn{3}{c}{Expect similar to 
Davis Campus} \\
Shotcrete Surface Coating        & \multicolumn{3}{c}{Under 
investigation} \\
\br
\end{tabular}
\end{center}
\end{table}

Long-term underground radon data have been collected since 2009 and at 
the Davis Campus since July 2012, shortly after outfitting was completed.  
Aside from the normal facility ventilation that brings fresh air 
underground via the Ross and Yates shafts, no extraordinary steps have 
been taken to mitigate underground radon levels.  Some research groups 
employ a type of purging system to reduce radon locally for their 
equipment.  As seen at other underground laboratories, a seasonal 
dependence is becoming apparent in the Davis Campus trends, with higher 
radon levels observed during the summer months May through September.  
The total average radon concentration over the monitoring period (883 
days) at the Davis Campus is approximately 310 Bq/m$^{3}$ with a low 
baseline of 150 Bq/m$^{3}$ during the winter months.  Brief excursions 
above 1000 Bq/m$^{3}$ have been observed, typically correlated with known 
events such as ventilation fan maintenance or when the Yates and/or Ross 
shafts are partially covered.

Other efforts to characterize physics backgrounds (eg., muons, neutrons, 
gamma rays) in various underground areas were carried out by various 
research groups.  Historically, the differential muon flux was measured by 
collaborators on the Homestake chlorine experiment~\cite{Cherry}.  More 
recently, since the mine was reopened as a laboratory, other groups have 
gathered data over several years~\cite{Bkgd-Muon, Bkgd-Radon, Bkgd-Gamma}.  
Results from recent neutron flux measurements collected on the 4850L are 
expected in an upcoming publication as are results related to the cosmic 
ray muon flux measured at the Davis Campus.

\section{Current Science Program}

Science efforts that started in 2007 during re-entry into the facility 
have grown steadily over the years.  Building on the legacy of the Ray 
Davis chlorine solar-neutrino experiment~\cite{Davis} that began in 1965 
at the Homestake Mine, twenty-six research groups have conducted 
underground research programs at SURF.  Measurements have been made and 
samples collected from elevations ranging from surface to the 5000L, 
investigating topics in physics, geology, biology and engineering.  A 
total of eighteen research programs are considered active as summarized in 
Table~\ref{tab:science_programs}, which includes a brief description of 
each group as well their respective locations.  The list includes the two 
main physics efforts underway at Davis Campus, as well as the two 
low-background counters.  There has also been significant interest from 
additional groups across all disciplines.


\begin{table}[htbp]
\caption{\label{tab:science_programs}Current scientific research programs 
at the Sanford Laboratory. Locations in bold indicate current 
installations or the subject of current focus; those in italic are 
imminent.}
\begin{center}
\begin{tabular}{llll}
\br
\multicolumn{1}{c}{\bf Experiment} & \multicolumn{1}{c}{\bf Description} & 
\multicolumn{1}{c}{\bf Locations}  & \multicolumn{1}{c}{\bf References} \\
\mr
\multicolumn{4}{l}{\bf Physics} \\
\mr
Large Underground Xenon & Dark matter using Xe & Surface, {\bf 
4850L} & {\scriptsize\cite{LUX-NIM,LUX}} \\
(LUX) &&&\\[0.08cm]
{\sc Majorana} & Neutrinoless double- & {\bf Surface}, 800L, & 
{\scriptsize\cite{MJD,MJD-Xu}} \\
{\sc Demonstrator} & beta decay using Ge & {\bf 4850L} & \\[0.08cm]
Center for Ultra-low Bkgd & Low-bkgd counter, & 
{\bf Surface}, 800L, & 
{\scriptsize\cite{Bkgd-Muon,Bkgd-Radon,Bkgd-Gamma,CUBED}} 
\\ Experiments in the Dakotas & isotope separation; & 
2000L, 4550L, & \\
(CUBED); also Bkgd & also bkgds: muon, & {\bf 4850L} &\\
Characterization & neutron, gamma, radon &&\\[0.08cm]
Berkeley Low-Bkgd Facility & Low-bkgd counter & {\bf 4850L} & 
{\scriptsize\cite{BLBF}} \\[0.08cm]
Compact Accelerator System & Neutron bkgds & 4100L, 4850L & 
\scriptsize\cite{CASPAR_neutron} \\
for Performing Astrophysical &&& \\
Research (CASPAR) &&&\\[0.08cm]
Long-Baseline Neutrino & Cleanliness tests & Surface, 4850L & 
{\scriptsize\cite{LBNE-Tiedt}} \\
Facility (LBNF) &&&\\[0.08cm]
Deep Underground Gravity & Seismic characterization
& {\bf Surface}, {\bf 4850L}, & {\scriptsize\cite 
{DUGL-2010_1,DUGL-2010_2}}\\
Laboratory (DUGL) & for gravity-wave & 300L, 800L, 1700L, &\\
& research & 2000L, 4100L &\\[-0.07cm]
\mr
\multicolumn{4}{l}{\bf Geology} \\
\mr
Geoscience Optical & Optical fiber & {\bf 2000L}, {\bf 4100L}, & 
{\scriptsize\cite{Geo-GEOXTM-2013,Geo-GEOXTM-2012_1,Geo-GEOXTM-2012_2,Geo-GEOXTM-2011},} \\
Extensometers and & applications, & {\bf 4850L} & 
{\scriptsize\cite{Geo-GEOXTM-2010}}\\
Tiltmeters (GEOX$^{\rm TM}$) & tiltmeters &&\\[0.08cm]
USGS Hydrogravity & Local gravity for water & 
{\bf Surface}, {\bf 300L}, & {\scriptsize\cite{Hydrogravity}} \\
& tables, densities & {\bf 4100L}, {\bf 4850L} &\\[0.08cm]
Petrology, Ore Deposits, & Core archive and logs, & 
Surface, 800L & 
{\scriptsize\cite{PODS-Steadman,PODS-Armstrong,PODS-Hamer}} \\
Structure (PODS) & geologic mapping && \\[0.08cm]
Transparent Earth & Seismic monitoring/ & 2000L, {\bf 4100L} & 
{\scriptsize\cite{Geo-TransparentEarth-2013,Geo-TransparentEarth-2011,Geo-TransparentEarth-2009}}\\ 
& properties, melt studies& {\bf 4850L} &\\[-0.07cm]
\mr
\multicolumn{1}{l}{\bf Biology} & \multicolumn{3}{l}{\footnotesize 
(Samples collected informally have also resulted in publications~\cite{Rinke-2013})} \\
\mr
Biodiversity (BHSU) & Microbiology & Surface, 300L, & 
{\scriptsize\cite{Bergmann-2014,Bergmann-2013,Bio-Waddell-2010}}\\
&& 2000L, {\bf 4100L}, & \\
&& 4550L, {\bf 4850L} &\\[0.08cm]
Biofuels (SDSMT) & Biofuels & 4550L, {\bf 4850L}, & 
{\scriptsize\cite{Bio-Bhalla-2013_1,Bio-Rastogi-2013_1,Bio-Bhalla-2013_2,Bio-Rastogi-2013_2},} \\
&& 5000L & 
{\scriptsize\cite{Bio-Rastogi-2010_1,Bio-Rastogi-2010_2,Bio-Rastogi-2009_1,Bio-Rastogi-2009_2}} \\[0.08cm] 
Lignocellulose (SDSU) & Biofuels & 1700L, {\bf 4850L} & 
{\scriptsize\cite{Bio-Bleakley-2011,Bio-Bleakley-2010}} \\[0.08cm] 
Syngas (SDSMT) & Biofuels & {\bf 4850L} & 
{\scriptsize\cite{Bio-Shende-2012_1,Bio-Shende-2012_2,Bio-Shende-2012_3}} 
\\[0.08cm]
Life Underground -- & Water in drill holes, & Surface, {\bf 800L}, & 
{\scriptsize\cite{NAI-2014}} \\
NASA Astrobiology Institute & geomicrobiology & {\bf 4850L}&\\[-0.07cm]
\mr
\multicolumn{4}{l}{\bf Engineering} \\
\mr
Xilinx, Inc. & Chip error testing & {\it 4850L} & \\[-0.07cm]
\br
\end{tabular}
\end{center}
\end{table}

\subsection{Large Underground Xenon (LUX)}

The LUX experiment~\cite{LUX-NIM} is conducting a direct search for 
weakly interacting massive particles (WIMPs) using 370~kg (118~kg 
fiducial) of purified xenon inside an ultra-pure titanium cryostat that 
is immersed in a water tank containing 272,550~liters (72,000~gallons) of 
purified water.  Following collisions in the liquid xenon volume, a 
strong electric field moves ionization electrons into the gas space at 
the top of the detector, and a total of 122 PMTs collect scintillation 
light from interactions in both xenon volumes.

Members of the LUX collaboration have been active onsite at SURF since 
December 2009, when preparations began for detector assembly at the 
Surface Laboratory.  After completing detector assembly and operational 
testing on surface, LUX began their transition underground to the Davis 
Campus in May 2012.  With strong support from SURF personnel, the 
fully-assembled LUX detector was successfully transported underground from 
the Surface Laboratory in the Yates cage and into the Davis Cavern -- 
stringent limits on acceleration and tilt were not exceeded.  The LUX 
space at the Davis Campus includes a cleanroom and a control room, which 
are separate from the main lab space.  The SURF water purification system 
and a storage area for liquid nitrogen are located near the LUX equipment 
but outside the clean environment.

The LUX collaboration published results in October 2013 from their first 
underground data run~\cite{LUX} and started a longer run (nominally 300 
live days) in October 2014. LUX is expecting to complete their 
experimental program in 2016.

\subsection{MAJORANA DEMONSTRATOR (MJD)}

The {\sc Majorana} collaboration is investigating neutrinoless 
double-beta decay using a detector called the {\sc Demonstrator} 
consisting 40 kg of germanium in two ultra-pure copper cryostats and a 
compact shield constructed from lead, copper and high-density 
polyethylene.  {\sc Majorana} expects to use up to 30~kg of enriched 
$^{76}$Ge.  The main technical goal of the {\sc Demonstrator} is to 
confirm that the ambitious purity requirements for a tonne-scale detector 
are achievable~\cite{MJD, MJD-Xu}.  The group also plans to test a 
controversial claim for the detection of neutrinoless double-beta decay.

In order to satisfy the very low background requirements, the {\sc 
Majorana} collaboration has set up a laboratory at the 4850L Ross Campus 
to produce the world's purest electroformed copper, which is then 
machined at a dedicated machine shop at the Davis Campus.  The MJD 
electroforming process begins with minimally-processed oxygen-free high 
conductivity copper nuggets (99.999\%), with typical properties: U = 
1.25~$\mu$Bq/kg, Th = 1.10~$\mu$Bq/kg and yield strength of $\sim$50~MPa.  
MJD electroformed copper averages U = 0.17~$\mu$Bq/kg, Th = 
0.06~$\mu$Bq/kg with yield strength 70--100~MPa (work hardened).  The 
average growth rate for electroformed copper is relatively 
slow at approximately 1~mm/month, and MJD expects to produce roughly 
2860~kg of the material for the {\sc Demonstrator}.

The {\sc Majorana} collaboration began work onsite starting in November 
2010 with deliveries of equipment (including the first germanium 
detectors) and then with preparations for the electroforming cleanroom.  
Production of electroformed copper began in July 2011 and about eight 
months later in March 2012 the group started to move equipment into the 
Davis Campus.  The majority of the detector shield is complete, including 
3400 cleaned and stacked Pb bricks and some of the inner copper sheets. 
MJD is currently collecting background data from germanium crystals in a 
prototype (commercial-copper) cryostat.  Completed assembly and the start 
of commissioning for the first production module with 20 kg of 
high-purity germanium (mostly enriched in $^{76}$Ge) is expected in early 
2015, with the second production module expected to be completed in late 
2015.  The full {\sc Majorana} experimental program with both production 
cryostats could extend into 2019.

\subsection{Center for Ultra-low Background Experiments in the Dakotas (CUBED)}

Initially created as a South Dakota Governor's Research Center 
administered through the University of South Dakota, CUBED involves 
physics researchers and others from the majority of universities in the 
state~\cite{CUBED}.  In addition to increasing the South Dakota academic 
involvement in SURF research, the center maintained a focus on areas of 
interest congruent with planned SURF experiments, such as crystal growth, 
low-background counting and stable isotope separation.

CUBED has set up a low-background counting laboratory in a dedicated room 
in the Lower Davis Cavern of the 4850L Davis Campus.  The system allows 
direct-gamma counting using a 1.3-kg high-purity n-type germanium crystal 
(60\% relative efficiency), resulting in 0.06--0.1~ppb sensitivities to 
U/Th as summarized in Table~\ref{tab:lbc_sensitivities}.  Installation of 
the germanium counter began in April 2013 and was finished a year later 
in April 2014 when the complete shield was established: 72 oxygen-free 
high conductivity copper bricks, a stainless steel enclosure for purging 
radon using boil-off nitrogen and approximately 500 lead bricks that were 
cleaned and coated in borated paraffin. Assays of community samples began 
in October 2014.  A subsequent reconfiguration of the detector shield 
carried out in December 2014 has resulted in a further 25\% reduction in 
the rate of background events compared to the original performance.  The 
CUBED low-background counter is expected to move to a separate dedicated 
cleanroom at the 4850L Ross Campus in Summer 2015.

Another CUBED project, the isotopic separation and ultra-purification 
(ISUP) project, is underway at the Surface Laboratory, where the 
separation of stable isotopes is being investigated.  Initial tests are 
being performed with CO$_{2}$, with plans to also study N$_{2}$, Ar, Kr 
and Xe.  Installation began in May 2014 with commissioning and full 
operation in September 2014.  Studies are expected to be completed in May 
2015.

A main CUBED focus involves crystal growth, and SURF personnel have 
provided a cost estimate to develop a laboratory to support underground 
crystal growth.  Funding is being sought.

\subsection{Berkeley Low-Background Facility (BLBF)}

After operating since the 1980s, LBNL counting operations at the Oroville 
Dam\footnote{The Oroville Dam was the original site of the UCSB/LBL 
double-beta decay experiment.} ended in February 2014.  The site was 
decommissioned and the HPGe detector and shield materials were shipped to 
SURF in March.  Starting in May, the detector and shielding were 
installed at a dedicated laboratory within the 4850L Davis Campus, next 
door to the CUBED low-background counter.  After a brief period of 
commissioning and background testing in June, the counter was back online 
for receiving samples in July~\cite{BLBF}.  In total, the implementation 
at SURF took less than 6 months.  Since July, eleven community samples 
have been counted on behalf of the {\sc Majorana} and LZ collaborations.

The BLBF counter is a 2.1-kg p-type coaxial germanium crystal (85\% 
relative efficiency), resulting in 0.01--0.3~ppb sensitivities to U/Th 
(see Table~\ref{tab:lbc_sensitivities}).  Compared to the original 
Oroville installation, background rates measured at SURF indicate a 
performance improvement of approximately 30\%, mainly due to shield 
configuration improvements.  The BLBF low-background counter is expected 
to move from the Davis Campus to a separate dedicated cleanroom at the 
4850L Ross Campus in Summer 2015.

\begin{table}[htbp]
\caption{\label{tab:lbc_sensitivities} Low-background counter 
sensitivities for a sample of order $\sim$1 kg and counting for 
approximately two weeks.}
\begin{center}
\begin{tabular}{lcccccc}
\br
              &  \multicolumn{2}{c}{\bf Uranium} &  \multicolumn{2}{c}{\bf 
Thorium}  &  \multicolumn{2}{c}{\bf Potassium} \\  
              & {\bf (ppt)} & {\bf (mBq/kg)} & {\bf (ppt)} & {\bf 
(mBq/kg)} & {\bf (ppb)} & {\bf (mBq/kg)} \\
\mr
CUBED         & 60 & 0.7    & 100 & 0.4    & 25 & 0.7  \\
BLBF          & 10 & 0.1    & 30  & 0.1    & 20 & 0.6  \\
\br
\end{tabular}
\end{center}
\end{table}

\subsection{Experiment Implementation Process}

SURF has developed a formal process for implementing 
experiments~\cite{SURF_science}.  Requirements include an experiment 
planning statement, a memorandum of understanding, appropriate insurance 
as well as a decommissioning plan.  If necessary, financial 
responsibilities for both SURF and the experiment are documented in a 
general services agreement.  Safety reviews are commensurate with the 
scale of the project.  SURF personnel are available to serve in various 
roles, including acting as a guide in the underground environment, 
providing logistical support as well as participating in the work planning 
process.

\section{Future Science}

\begin{figure}
\begin{center}
\includegraphics*[scale=0.157]{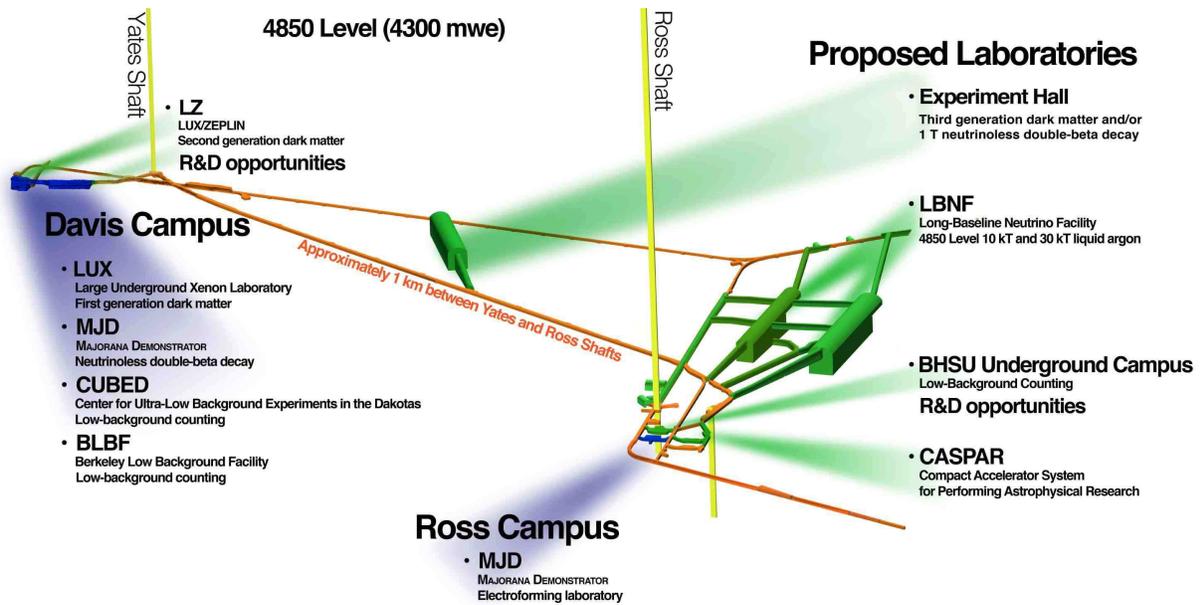}    
\caption{\label{fig:4850LCurrentFuture} The 4850L of SURF highlighting the 
existing and proposed experiments.}
\end{center}
\end{figure}

The future offers many scientific opportunities for underground science 
that can be pursued at SURF.  Both the nuclear physics and high-energy 
physics communities are developing strategic plans for the future of 
their respective fields, and projects that are planned for SURF figure 
prominently~\cite{P5}.  Near-term expansions are underway and there exist 
options to significantly expand the facility footprint to accommodate 
additional endeavors.  The locations of existing and future experiments 
are shown in Figure~\ref{fig:4850LCurrentFuture}.

\subsection{Black Hills State University Underground Campus (BHUC)}

\begin{figure}
\begin{center}
\includegraphics*[scale=0.58]{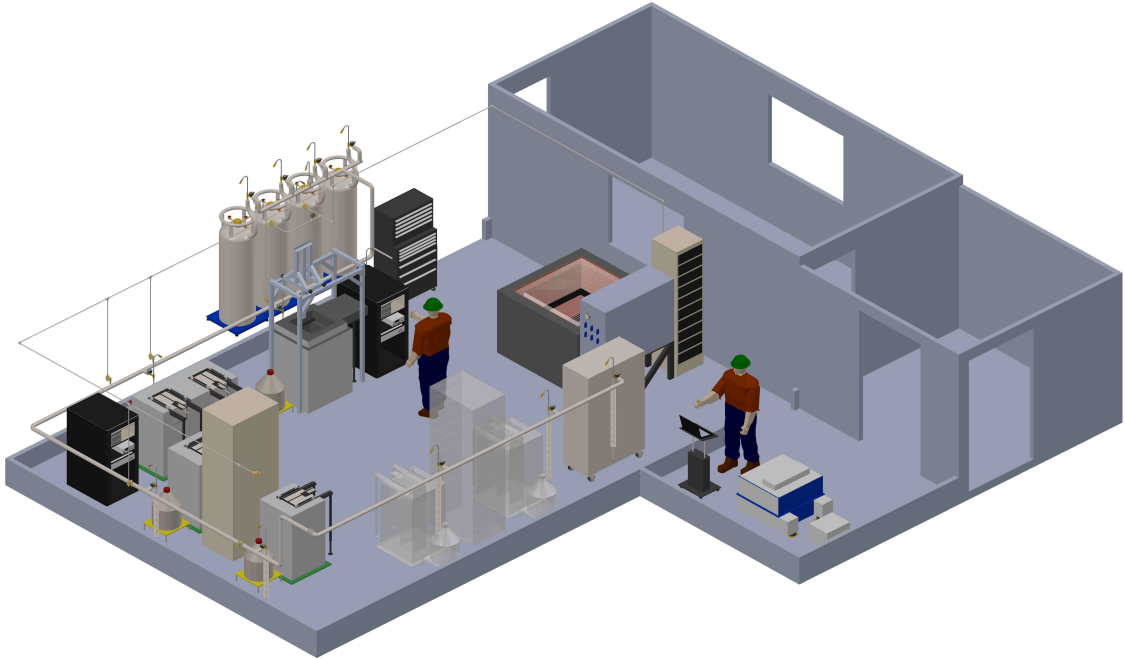}
\caption{\label{fig:BHUC_layout}Schematic of the Black Hills State 
University Underground Campus at the 4850L Ross Campus, showing a 
nominal set of low-background assay instruments and a liquid nitrogen 
supply and purge system.  The rooms on the right-hand side are intended 
to be used to conduct biology and geology research.}
\end{center}
\end{figure}

As mentioned previously, Black Hills State University is developing an 
underground campus at the 4850L Ross Campus, including a cleanroom that 
will host a variety of research activities.  Approximately 75\% of the 
BHUC cleanroom space is envisioned to be a low-background assay 
laboratory operated at Class 1000, while the remaining 25\% of the 
laboratory will be reserved for research that does not require a high 
level of cleanliness.  Space in the cleanest laboratory area dedicated to 
performing low-background assays is expected to house 8--9 instruments, 
possibly including more than a half-dozen high-purity germanium detectors 
and the associated shielding and access systems as shown in 
Figure~\ref{fig:BHUC_layout}.  The area outside the cleanroom will be 
used for storage as well as other research activities.

\subsection{Compact Accelerator System for Performing Astrophysical 
Research (CASPAR)}

An underground facility is under development at the 4850L Ross Campus to 
study the stellar nuclear fusion reactions responsible for the production 
of elements heavier than iron, especially via the slow neutron-capture 
nucleosythesis process (s-process).  The Compact Accelerator System for 
Performing Astrophysical Research (CASPAR) collaboration seeks to study 
($\alpha$,n) reactions at stellar energies relevant for helium and carbon 
burning~\cite{CASPAR_reactions}.  CASPAR will use an existing 1-MV Van de 
Graff accelerator from the University of Notre Dame capable of producing 
high-intensity ($\sim$150~$\mu$A) proton and helium beams with energies 
between 200~keV and 1~MeV.  A specialized windowless gas target system for 
$^{22}$Ne and $^{13}$C is also being developed at the Colorado School of 
Mines.  The South Dakota School of Mines and Technology will operate the 
facility.

The initial CASPAR scientific program is expected to take 5--8 years to 
complete, with a broader program that is expected to last for over a 
decade.  Potential future federal funding could allow for expansion to 
include additional elements that were originally planned for the Dual Ion 
Accelerators for Nuclear Astrophysics (DIANA) project~\cite{DIANA}.

\subsection{LUX-ZEPLIN (LZ)}

\begin{figure}
\begin{center}
\includegraphics*[scale=0.46]{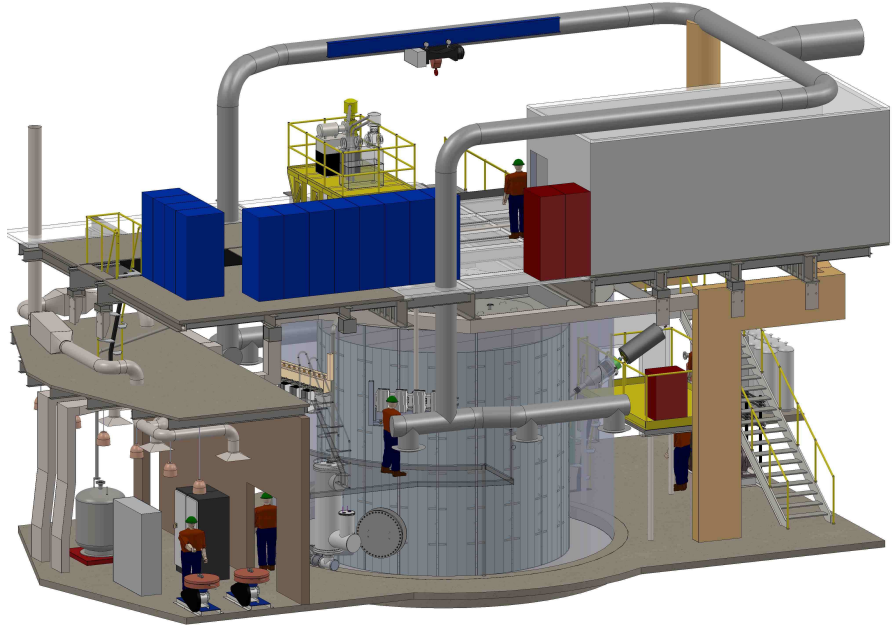} 
\caption{\label{fig:LZ_layout}Schematic of the LUX-ZEPLIN experiment at 
the 4850L Davis Campus.  Existing infrastructure is shown such as the 
ductwork, the water shielding tank (lower center) and the control room 
(upper right).  LZ will require all available space in the Davis Cavern, 
including rooms (lower left) currently used by the two low-background 
counters.}
\end{center}
\end{figure}

In July 2014, the LUX-ZEPLIN (LZ) proposal was one of two direct-search 
generation-2 dark matter experiments selected for funding by the U.S.\ DOE 
division of High Energy Physics.  The LZ detector will employ 
approximately 10~tonnes of liquid xenon ($\sim$6 tonnes fiducial or 
$\sim$45$\times$ LUX fiducial), and will rely on modest upgrades to 
existing SURF infrastructure, including the Surface Laboratory for 
detector assembly and underground space at the 4850L Davis Campus where 
the LUX experiment is currently operating.  In particular, no new 
excavation will be required, and LZ will be able to reuse the water 
shielding tank used for LUX with the addition of an improved background 
veto system consisting of gadolinium-loaded linear alkyl-benzene liquid 
scintillator, as shown in Figure~\ref{fig:LZ_layout}.  LZ sensitivity is 
projected to be approximately 100$\times$ better than the anticipated LUX 
300-day result (underway), with sensitivity of 
2$\times$10$^{-48}$~cm$^{2}$ for its full 1000-day exposure.

Detector assembly at the Surface Laboratory could begin as early as 
mid-2015. Underground production data-taking is scheduled to begin 2019 
and will nominally last 5 years.

\subsection{Neutrinoless Double-Beta Decay}

The {\sc Majorana} collaboration is investigating performing the 
next-generation neutrinoless double-beta decay experiment on the 4850L at 
SURF.  If the combination of rock overburden and veto is determined to be 
sufficient, scaling up to order $\sim$200~kg from the current {\sc 
Demonstrator} may be possible at the Davis Campus but ultimately a new 
laboratory may be needed.

\subsection{Long Baseline Neutrino Facility (LBNF)}

Neutrinos propagated over long distances ($>$ 1000~km) allow for 
investigations into fundamental physics, touching on such compelling 
unanswered puzzles as neutrino mass hierarchy and CP violation 
(matter/antimatter asymmetry) as well as many other topics.

For many years, the U.S.\ has been developing an experiment called the 
Long Baseline Neutrino Experiment (LBNE)~\cite{LBNE} that is transforming 
into the Long Baseline Neutrino Facility (LBNF).  While the project design 
has undergone many iterations (consideration of water \v{C}erenkov 
technology and an initial detector installation on the surface), the 
latest design consists of a large liquid argon detector (or detectors) 
located underground at the 4850L of SURF that would observe neutrinos 
generated approximately 1300~km away at Fermilab using an upgraded 
accelerator beam (up to 2.4~MW).  The detector mass could be staged in 
various ways (e.g., 10~kt, 30~kt, etc.) to reach a nominal fiducial mass 
of 40~ktonne (total mass $\sim$60~ktonne).  Some simulations to quantify 
backgrounds due to cosmic ray muons have already been 
performed~\cite{Bkgd-Muon-LBNE}.

International partners are required to ensure the success of such an 
ambitious project, and a process is currently underway to consolidate 
interested researchers into a single Fermilab-led experiment.  The new 
proto-collaboration is referred to as the Experiment at the Long Baseline 
Neutrino Facility (ELBNF) and is expected to include the majority of 
researchers from existing collaborations such as LBNE (90 institutions 
and 500+ collaborators) and the Long Baseline Neutrino Oscillation (LBNO) 
European experiment~\cite{LBNO}.  A Letter of Intent defining the new 
collaboration has been drafted~\cite{ELBNF_LOI} and will be presented to 
the Fermilab Program Advisory Committee (PAC) in January 2015.

Geotechnical studies have already commenced, and the first construction 
efforts associated with LBNF could begin as early as FY16 when shaft 
services and some aspects of a waste-rock handling system could be 
installed.  Underground excavation could begin as soon as the Ross shaft 
rehabilitation is completed, expected to be mid-2017.

\subsection{Future Expansion}

Space is currently available for modest R\&D efforts, including cutouts at 
the Davis Campus (30--33 m$^{2}$) and laboratory space at the BHSU 
Underground Campus. While it may be possible to utilize a few other 
existing excavations on the 4850L and other relatively deep levels, most 
practical existing excavations will soon be occupied.  Some spaces that 
are currently occupied may become available again in 5--10 years.

New excavations of order 11,500 cubic meters (15,000 cubic yards) of rock 
could be performed prior to the completion of the Ross Shaft 
rehabilitation; however, once a waste-rock handling system is installed to 
support the LBNF project very large excavation projections become 
possible.  Based on the DUSEL preliminary design, a generic Experiment 
Hall is included on Figure~\ref{fig:4850LCurrentFuture} as a placeholder 
to accommodate a tonne-scale neutrinoless double-beta decay experiment 
and/or a generation-3 dark matter experiment.  Nominal dimensions are 20~m 
W $\times$ 24~m H $\times$ 50--100~m L displacing 45,900--85,600 cubic 
meters (60,000--112,000 cubic yards) of rock, including access tunnels, 
over-break and rock swell.  Other configurations and locations are also 
possible.

\section{Summary}

SURF is a deep underground research facility, dedicated to scientific 
uses.  Research activities are supported at a number of facilities, both 
on the surface and underground.  Two campuses on the 4850L accommodate a 
number of leading efforts, and in particular the 4850L Davis Campus has 
been successfully operating for over 2.5~years.  The LUX and {\sc 
Majorana} experiments are well established and there are current 
capabilities for low-background counting.  Many expansion possibilities 
are on the horizon and a number of key experiments in the U.S.\ research 
program are developing plans for installation at SURF.


\section*{References}
\bibliography{Heise}

\providecommand{\newblock}{}
\begin{thebibliography}{10}
\expandafter\ifx\csname url\endcsname\relax
  \def\url#1{{\tt #1}}\fi
\expandafter\ifx\csname urlprefix\endcsname\relax\def\urlprefix{URL }\fi
\providecommand{\eprint}[2][]{\url{#2}}

\bibitem{SURF-Heise-2014}
Heise J 2014 {\em AIP Conf. Proc.\/} {\bf 1604} 331--344 (\textit{Preprint}
  \eprint{arXiv:1401.0861 [physics.ins-det]})

\bibitem{SURF-Lesko}
Lesko K~T 2012 {\em Eur. Phys. J. Plus\/} {\bf 127} 107--118

\bibitem{SURF-Heise-2010}
Heise J 2010 {\em Nuc. Phys. A\/} {\bf 834} 805c--808c

\bibitem{Mitchell-2009}
Mitchell S~T 2009 {\em Nuggets to Neutrinos: The Homestake Story\/} (Xlibris)

\bibitem{DUSEL-Lesko}
Lesko K~T 2013 Why the {US} needs a deep domestic research facility: Owning
  rather than renting the education benefits, technology advances, and
  scientific leadership of underground physics (\textit{Preprint}
  \eprint{arXiv:1304.0402 [physics.ins-det]})

\bibitem{DUSEL_PDR-Lesko}
Lesko K~T {\em et~al.\/} 2012 {D}eep {U}nderground {S}cience and {E}ngineering
  {L}aboratory --- preliminary design report (\textit{Preprint}
  \eprint{arXiv:1108.0959 [hep-ex]})

\bibitem{Murdoch-2012}
Murdoch L~C, Germanovich L~N, Wang H, Onstott T~C, Elsworth D, Stetler L and
  Boutt D 2012 {\em Hydrogeol. J.\/} {\bf 20} 27--43

\bibitem{Zhan_Duex-2010}
Zhan G and Duex T 2010 {\em Mining Eng.\/} {\bf 64(4)} 64--68

\bibitem{BHUC}
Mount B 2015 Low background counting facility in {BHSU} {U}nderground {C}ampus
  at {S}anford {L}ab {\it {P}roceedings for this workshop}

\bibitem{Pariseau-2012}
Pariseau W~G, Tesarik D~R and Trancynger T~C 2012 {\em Transactions of the
  Society for Mining, Metallurgy, and Exploration\/} {\bf 332} 370--388

\bibitem{BLBF_report}
Thomas K~J, Lesko K~T and Smith A~R 2014 Berkeley low background facility
  update {\it Private Communication}

\bibitem{Bkgd-Muon}
Gray F~E, Ruybal C, Totushek J, Mei D~{\mbox{-}M}, Thomas K and Zhang C 2010
  {\em Nucl. Instrum. Meth. A\/} {\bf 638} 63--66 (\textit{Preprint}
  \eprint{arXiv:1007.1921 [nucl-ex]})

\bibitem{RoggenthenHart-2014}
Hart K, Trancynger T~C, Roggenthen W and Heise J 2014 {\em Proceedings of the
  South Dakota Academy of Science\/} {\bf 93} 33--41
  {\url{http://www.sdaos.org/wp-content/uploads/pdfs/2014/roggenthen\%2033-41.%
pdf}}

\bibitem{Caddey-1991}
Caddey S~W, Bachman R~L, Campbell T~J, Reid R~R and Otto R~P 1991 {\em U.S.
  Geol. Surv. Bull.\/} {\bf 1857-J} J1--J67

\bibitem{Assay-Whitepaper}
Roggenthen W and Smith A~R 2008 White paper: {U}, {T}h, {K} contents of
  materials associated with the {H}omestake {DUSEL} site, {L}ead, {S}outh
  {D}akota {\it Private Communication}

\bibitem{Assay-Oroville}
Smith A~R 2007 and updates through 2014 {H}omestake samples: Results of
  radiometric analyses at {LBNL} {\it Private Communication}

\bibitem{Assay-YDC}
Chan Y~D 2012 The low-background construction of laboratories at the 4850-ft
  level {D}avis {C}ampus {\tt http://www.sanfordlab.org/lbnl/186}

\bibitem{Cherry}
Cherry M~L, Deakyne M, Lande K, Lee C~K, Steinberg R~I, Cleveland B and Fenyves
  E~J 1983 {\em Phys. Rev. D\/} {\bf 27} 1444--1447

\bibitem{Bkgd-Radon}
Thomas K~J 2011 {\em Radon monitoring and the role of iron oxide on radon
  emanation at the {S}anford {L}aboratory at {H}omestake in {L}ead, {S}outh
  {D}akota\/} Master's thesis (unpublished) University of South Dakota
  Department of Physics

\bibitem{Bkgd-Gamma}
Mei D~{\mbox{-}M}, Zhang C, Thomas K and Gray F 2010 {\em Astropart. Phys.\/}
  {\bf 34} 33--39 (\textit{Preprint} \eprint{arXiv:0912.0211 [nucl-ex]})

\bibitem{Davis}
Cleveland B~T, Daily T, R~Davis J, Distel J~R, Lande K, Lee C~K, Wildenhain P~S
  and Ullman J 1998 {\em Astrophys. J.\/} {\bf 496} 505--526

\bibitem{LUX-NIM}
Akerib D~S {\em et~al.\/} 2013 {\em Nucl. Instrum. Meth. A\/} {\bf 704}
  111--126 (\textit{Preprint} \eprint{arXiv:1211.3788 [physics.ins-det]})

\bibitem{LUX}
Akerib D~S {\em et~al.\/} 2014 {\em Phys. Rev. Lett.\/} {\bf 112} 091303
  (\textit{Preprint} \eprint{arXiv:1310.8214 [astro-ph.CO]})

\bibitem{MJD}
Abgrall N {\em et~al.\/} 2014 {\em Advances in High Energy Physics\/} {\bf
  2014} Article ID 365432, 18 pp (\textit{Preprint} \eprint{arXiv:1308.1633
  [physics.ins-det]})

\bibitem{MJD-Xu}
Xu W {\em et~al.\/} 2015 The {\sc {m}ajorana {d}emonstrator}: A search for
  neutrinoless double-beta decay of $^{76}${G}e {\it {P}roceedings for this
  workshop} (\textit{Preprint} \eprint{arXiv:1501.03089 [nucl-ex]})

\bibitem{CUBED}
Szczerbinska B {\em et~al.\/} 2010 {\em Proceedings of the South Dakota Academy
  of Science\/} {\bf 89} 23

\bibitem{BLBF}
Berkeley low background facility {\url{http://lbf.lbl.gov}}

\bibitem{CASPAR_neutron}
Best A, G{\"o}rres J, Junker M, Kratz K~{\mbox{-}L}, Long A, Smith K and
  Wiescher M Low energy neutron background in deep underground laboratories
  {T}o be submitted to {\it Astropart. Phys.} in 2015

\bibitem{LBNE-Tiedt}
Tiedt D 2013 {\em Radioactive background simulation and cleanliness standards
  analysis for the {L}ong {B}aseline {N}eutrino {E}xperiment located at the
  {S}anford {U}nderground {R}esearch {F}acility\/} Master's thesis
  (unpublished) South Dakota School of Mines and Technology Department of
  Physics

\bibitem{DUGL-2010_1}
Harms J {\em et~al.\/} 2010 {\em Class. Quant. Grav.\/} {\bf 27} 225011
  (\textit{Preprint} \eprint{arXiv:1006.0678 [gr-qc]})

\bibitem{DUGL-2010_2}
Acernese F, De~Rosa R, De~Salvo R, Garufi F, Giordano G, Harms J, Mandic V,
  Sajeva A, Trancynger T and Barone F 2010 {\em J. Phys.: Conf. Ser\/} {\bf
  228} 012036

\bibitem{Geo-GEOXTM-2013}
Gage J~R, Fratta D, Turner A~L, MacLaughlin M~M and Wang H~F 2013 {\em Int. J.
  Rock Mech. Mining Sci.\/} {\bf 61} 244--255

\bibitem{Geo-GEOXTM-2012_1}
Volk J {\em et~al.\/} 2012 {\em J. Inst.\/} {\bf 7} P01004 (\textit{Preprint}
  \eprint{arXiv:1205.1777 [physics.acc-ph]})

\bibitem{Geo-GEOXTM-2012_2}
Volk J, Shiltsev V, Chupyra A, Kondaurov M, Singatulin S, Fratta D, Meulemans
  A, Potier C and Wang H 2012 {\em 12th International Workshops on Accelerator
  Alignment IWAA, 10--14 September, Fermilab\/}

\bibitem{Geo-GEOXTM-2011}
Wang H~F, Gage J~R, Fratta D~O, MacLaughlin M~M and Turner A 2011 {\em
  Proceedings of the 8th International Workshop on Structural Health
  Monitoring, 13--15 September, Stanford, CA\/}  8 pp

\bibitem{Geo-GEOXTM-2010}
Gage J~R, Noni N, Turner A, MacLaughlin M and Wang H~F 2010 {\em 44th U.S. Rock
  Mechanics Symposium and 5th U.S-Canada Rock Mechanics Symposium, 27--30 June,
  Salt Lake City, UT\/}  8 pp

\bibitem{Hydrogravity}
Kennedy J, Murdoch L, Long A~J and Koth K 2011 {\em American Geophysical Union
  Fall Meeting, 5--9 December, San Francisco, CA\/} {\bf Abstract NH54A-01} A1

\bibitem{PODS-Steadman}
Steadman J 2015 {\em BIFs, black shales, and gold deposits: A reevaluation\/}
  Ph.{D}. thesis (unpublished) University of Tasmania School of Earth Sciences

\bibitem{PODS-Armstrong}
Armstrong A~E 2013 {\em Hydrothermal alteration and gold mineralization in
  biotite zone host rocks at {H}omestake {M}ine in {L}ead, {S}outh {D}akota\/}
  Master's thesis (unpublished) South Dakota School of Mines and Technology
  Department of Geology

\bibitem{PODS-Hamer}
Hamer R~C 2010 {\em Hydrothermal alteration and gold deposition in the
  {H}omestake iron-formation-hosted gold deposit, {L}ead, {S}outh {D}akota\/}
  Master's thesis (unpublished) South Dakota School of Mines and Technology
  Department of Geology

\bibitem{Geo-TransparentEarth-2013}
Roggenthen W~M and Koch C~D 2013 {\em Proc. 47th U.S. Rock Mech./Geomech. Symp.
  23--26 June, 2013, San Francisco, CA\/} {\bf Paper 13-493} 8

\bibitem{Geo-TransparentEarth-2011}
Sherman C~S, Magliocco M and Glaser S~D 2011 {\em 45th U.S. Rock Mech./Geomech.
  Symp., 26--29 June, San Francisco, CA\/} {\bf Paper 11-526} 7

\bibitem{Geo-TransparentEarth-2009}
Van~Beek J~K, Roggenthen W~M, Magliocco M and Glaser S~D 2009 {\em Proceedings
  of the 3rd CANUS Rock Mechanics Symposium, May 2009, Toronto, ON\/} {\bf
  Paper 4155} 7

\bibitem{Rinke-2013}
Rinke C {\em et~al.\/} 2013 {\em Nature\/} {\bf 499} 431--437

\bibitem{Bergmann-2014}
Bergmann D~J, Anderson C~M and Gorbatenko O 2014 {\em Proceedings of the South
  Dakota Academy of Science\/} {\bf 93} 214

\bibitem{Bergmann-2013}
Finch A~E, Brakke A, Gorbatenko O, Anderson C~M and Bergmann D~J 2013 {\em
  Proceedings of the South Dakota Academy of Science\/} {\bf 92} 173

\bibitem{Bio-Waddell-2010}
Waddell E~J, Elliott T~J, Vahrenkamp J~M, Roggenthen W~M, Sani R~K, Anderson
  C~M and Bang S~S 2010 {\em Env. Tech.\/} {\bf 31:8--9} 979--991

\bibitem{Bio-Bhalla-2013_1}
Bhalla A, Kainth A and Sani R~K 2013 {\em Genome Announc.\/} {\bf 1} e00595--13

\bibitem{Bio-Rastogi-2013_1}
Rastogi G, Gurram R~N, Bhalla A, Jaswal R, Gonzalez R, Bischoff K~M, Hughes
  S~R, Sudhir K and Sani R~K 2013 {\em Journal of Frontiers in
  Microbiotechnology, Ecotoxicology and Bioremediation\/} {\bf 4} 1--8

\bibitem{Bio-Bhalla-2013_2}
Bhalla A, Bansal N, Kumar S and Sani R~K 2013 {\em Bio. Tech.\/} {\bf 128}
  751--759

\bibitem{Bio-Rastogi-2013_2}
Rastogi G, Gurram R~N, Bhalla A, Gonzalez R, Bischoff K~M, Hughes S~R, Kumar S
  and Sani R~K 2013 {\em Front. Microbiol.\/} {\bf 4} 18

\bibitem{Bio-Rastogi-2010_1}
Rastogi G, Bhalla A, Adhikari A, Bischoff K~M, Hughes S~R, Christopher L~P and
  Sani R~K 2010 {\em Bio. Tech.\/} {\bf 101} 8798--8806

\bibitem{Bio-Rastogi-2010_2}
Rastogi G, Osman S, Kukkadapu R~K, Engelhard M, Vaishampayan P~A, Andersen G~L
  and Sani R~K 2010 {\em Microb. Ecol.\/} {\bf 60} 539--550

\bibitem{Bio-Rastogi-2009_1}
Rastogi G, Muppidi G~L, Gurram R~N, Adhikari A, Bischoff K~M, Hughes S~R, Apel
  W~A, Bang S~S, Dixon D~J and Sani R~K 2009 {\em J. Ind. Microbiol.
  Biotechnol.\/} {\bf 36} 585--598

\bibitem{Bio-Rastogi-2009_2}
Rastogi G, Stetler L~D, Peyton B~M and Sani R~K 2009 {\em J. Microb.\/} {\bf
  47} 371--384

\bibitem{Bio-Bleakley-2011}
Murthy N~K~S, Bleakley B~H and Speidel R 2011 {\em Proc. 61st Society for
  Industrial Microbiology Annual Meeting, 24--28 July, New Orleans, LA\/} {\bf
  Abstract P99}

\bibitem{Bio-Bleakley-2010}
Harris S~T, Gibson K~A and Bleakley B~H 2010 {\em Proc. 60th Society for
  Industrial Microbiology Annual Meeting, 1--5 August, San Francisco, CA\/}
  {\bf Abstract P35}

\bibitem{Bio-Shende-2012_1}
Shende A, Harder-Heinze D and Shende R 2012 {\em Proceedings of the South
  Dakota Academy of Science\/} {\bf 91} 208

\bibitem{Bio-Shende-2012_2}
Shende A, Jaswal R, Harder-Heinze D, Menan A and Shende R 2012 {\em Technical
  Proceedings of the 2012 NSTI Nanotechnology Conference and Expo\/} {\bf 3}
  489

\bibitem{Bio-Shende-2012_3}
Shende A, Jaswal R and Shende R 2012 {\em Proceedings of AIChE Annual
  Meeting\/} {\bf ISBN: 978-1-4665-6277-6} 120

\bibitem{NAI-2014}
Osburn M~R, LaRowe D~E, Momper L~M and Amend J~P 2014 {\em Front. Microbiol.\/}
  {\bf 5} 610:1--14

\bibitem{SURF_science}
{SURF} science liaison office
  {\url{http://www.sanfordlab.org/science-liaison-office}}

\bibitem{P5}
 2014 Report of the particle physics project prioritization panel (p5)
  {\url{http://www.usparticlephysics.org/p5}}

\bibitem{CASPAR_reactions}
Best A {\em et~al.\/} 2013 {\em Phys. Rev. C\/} {\bf 87} 045805
  (\textit{Preprint} \eprint{arXiv:1304.6443 [nucl-ex]})

\bibitem{DIANA}
Best A, Couder M, Famiano M, Lemut A and Wiescher M 2013 {\em Nucl. Instrum.
  Meth. A\/} {\bf 727} 104--108 {\tt http://www.jinaweb.org/underground/DIANA}

\bibitem{LBNE}
Adams C {\em et~al.\/} 2013 Scientific opportunities with the {L}ong-{B}aseline
  {N}eutrino {E}xperiment (\textit{Preprint} \eprint{arXiv:1307.7335 [hep-ex]})

\bibitem{Bkgd-Muon-LBNE}
Barker D, Mei D~{\mbox{-}M} and Zhang C 2012 {\em Phys. Rev. D\/} {\bf 86}
  054001 (\textit{Preprint} \eprint{arXiv:1202.5000 [physics.ins-det]})

\bibitem{LBNO}
Agarwalla S~K {\em et~al.\/} 2014 {\em J. High Energy Phys.\/} {\bf 1405} 094
  (\textit{Preprint} \eprint{arXiv:1312.6520 [hep-ph]})

\bibitem{ELBNF_LOI}
An experimental program in neutrinos, nucleon decay and astroparticle physics
  enabled by the {F}ermilab {L}ong-{B}aseline {N}eutrino {F}acility
  {\url{https://indico.fnal.gov/getFile.py/access?resId=0\&materialId=0\&confI%
d=9214}} {L}etter of Intent submitted to the Fermilab PAC

\end{thebibliography}

\end{document}